\RequirePackage{fix-cm}
\documentclass[twocolumn]{svjour3}          
\smartqed  
\usepackage{graphicx}
\usepackage{amsmath}
\usepackage{amsfonts}
\usepackage{amssymb}
\usepackage{booktabs}
\usepackage{textcomp}
\usepackage{enumitem}
\usepackage[round]{natbib}
\usepackage{bm}
\usepackage{url}
\usepackage{tikz}
\usepackage{makecell}
\definecolor{gruen}{RGB}{255,140,0}

\usepackage{mathtools}

\DeclareMathOperator\erf{erf}
\newcommand{\beginappendix}{%
        \setcounter{section}{0}
        \renewcommand{\thesection}{A\arabic{section}}%
        \setcounter{table}{0}
        \renewcommand{\thetable}{A\arabic{table}}%
        \setcounter{figure}{0}
        \renewcommand{\thefigure}{A\arabic{figure}}%
        \setcounter{equation}{0}
        \renewcommand{\theequation}{A\arabic{equation}}
     }

\journalname{arxiv}

\setcitestyle{authoryear,open={(},close={)}} 

\begin{document}

\title{A computational framework for modeling cell-matrix interactions in soft biological tissues}
\titlerunning{Computational framework for modeling cell-matrix interactions in soft tissues} 

\author{Jonas F. Eichinger \and Maximilian J. Grill \and Iman Davoodi Kermani \and Roland C. Aydin \and Wolfgang A. Wall \and Jay D. Humphrey \and Christian J. Cyron}

\institute{
J. F. Eichinger $\cdot$ M. J. Grill $\cdot$ I. Davoodi Kermani $\cdot$ W. A. Wall
\at Institute for Computational Mechanics, Technical University of Munich,\\ 
Garching, 85748 Germany
\and
J. F. Eichinger $\cdot$ C. J. Cyron
\at Institute of Continuum and Materials Mechanics, Hamburg University of Technology,\\ 
Hamburg, 21073 Germany
\and 
J. D. Humphrey
\at Department of Biomedical Engineering, Yale University,\\
New Haven, CT 06520 USA
\and
R. C. Aydin $\cdot$ C. J. Cyron
\at Institute of Material Systems Modeling, Helmholtz-Zentrum Geesthacht,\\
Geesthacht, 21502 Germany\\
\email{christian.cyron@tuhh.edu}
}

\date{Received: date / Accepted: date}

\maketitle

\begin{abstract}

Living soft tissues appear to promote the development and maintenance of a preferred mechanical state within a defined tolerance around a so-called set-point. This phenomenon is often referred to as mechanical homeostasis. In contradiction to the prominent role of mechanical homeostasis in various (patho)physiolo\-gical processes, its underlying micromechanical mechanisms acting on the level of individual cells and fibers remain poorly understood, especially, how these mechanisms on the microscale lead to what we macroscopically call mechanical homeostasis. Here, we present a novel finite element based computational framework that is constructed bottom up, that is, it models key mechanobiological mechanisms such as actin cytoskeleton contraction and molecular clutch behavior of individual cells interacting with a reconstructed three-dimensional extracellular fiber matrix. The framework reproduces many experimental observations regarding mechanical homeostasis on short time scales (hours), in which the deposition and degradation of extracellular matrix can largely be neglected. This model can serve as a systematic tool for future \textit{in silico} studies of the origin of the numerous still unexplained experimental observations about mechanical homeostasis.

\end{abstract}

\begin{acknowledgements}
We gratefully thank Diane Tchibozo and Lisa Pretsch for their contributions to Fig. 7C. We thank Jonas Koban for his contributions to the development of the code that generated the fiber networks as input to our simulations.
\end{acknowledgements}

\vspace{0.5cm}

\noindent {\Large\textbf{Declarations}}

\noindent\textbf{Funding} \newline
Funded by the Deutsche Forschungsgemeinschaft (DFG, German Research Foundation) – Projektnummer 257981\-274, Projektnummer 386349077. The authors also gratefully acknowledge financial support by the International Graduate School of Science and Engineering (IGSSE) of Technical University of Munich, Germany. \\

\noindent\textbf{Conflicts of interest/Competing interests} \newline
Not applicable\\

\noindent\textbf{Availability of data and material} \newline
Not applicable\\

\noindent\textbf{Code availability} \newline
Not applicable

\newpage
\section{Introduction}
\label{sec:intro}

Living soft tissues, in contrast to classical engineering materials, usually seek to establish and maintain a mechanical state that is not stress-free. This behavior of living soft tissues is often referred to as \textit{mechanical homeostasis}, and it plays a key role in the control of form and function in health and disease \citep{Lu2011,Humphrey2014,Ross2013,Cox2011,Bonnans2014}. Intra-cellular structures such as the actomyosin cytoskeleton are physically coupled to the surrounding extracellular matrix (ECM) via transmembrane protein complexes such as integrins that can cluster to form focal adhesions \citep{Cavalcanti-Adam2007,Lerche2019}. This coupling allows cells to receive mechanical cues from their environment, transduce these cues into intra-cellular signals, and react, for example, by adapting cellular stress and thereby also the stress of the surrounding ECM. Physical interactions between cells and ECM have been shown to control various processes on the cellular scale such as cell migration \citep{Kim2019,Xie2017,Hall2016,Grinnell2010}, differentiation \citep{Chiquet2009,Mammoto2012,Zemel2015,Seo2020} and survival \citep{Bates1995,ZHU2002,Sukharev2012,Schwartz1995}, and are therefore fundamental for health and in disease of entire tissues and organs.

To study the micromechanical foundations of mechanical homeostasis experimentally, tissue culture studies with cell-seeded collagen or fibrin gels have attracted increasing interest over the past decades \citep{Eichinger2020b}. Circular free-floating gels, when seeded with fibroblasts, exhibit a strong compaction over multiple days in culture due to cellular contractile forces \citep{Simon2014,Simon2012}. Studies of such gels whose compaction is prevented by boundary constraints typically show a two-phase response. First, tension in the gels rapidly increases to a specific value, the so-called homeostatic tension (phase I), and then remains largely constant (phase II) for the rest of the experiment \citep{Marenzana2006,Brown1998,Ezra2010,Eichinger2020,Brown2002,Courderot-Masuyer2017,Campbell2003a,Dahlmann-Noor2007,Karamichos2007,Sethi2002}. If the gel is perturbed in phase II, for example, by an externally imposed deformation, cells appear to promote a restoration of the homeostatic state \citep{Brown1998,Ezra2010}. Despite substantial research efforts over decades, the exact interplay between cells and surrounding tissue that is crucial for mechanical homeostasis and other related phenomena such as durotaxis still remains poorly understood to date \citep{Eichinger2020b}.

Computational studies in this field have focused primarily on decellularized ECM systems to study the micromechancial and physical properties of networks of fibers  \citep{Heussinger2007,Mickel2008,Lindstrom2010,Chatterjee2010,Broedersz2011a,Stein2011,Ronceray2015,Cyron2012,Cyron2013,Cyron2013a,Lang2013,Motte2013,Muller2014,Jones2014,Lee2014,Muller2015,Mauri2016,Dong2017,Humphries2018,Zhou2018,Bircher2019,Domaschke2019,Domaschke2020}. Current computational models of cell-ECM interactions often suffer from shortcomings -- most are limited to two dimensions and just one or two cells \citep{Wang2014,Abhilash2014,Notbohm2015,Jones2015,Kim2017,Grimmer2018,Burkel2018,Humphries2017}. The importance of the third dimension for the physics of fiber networks is well-known \citep{Baker2012,Jansen2015,Duval2017}, and it can be assumed that collective interactions between more than just two cells play important roles in mechanical homeostasis.
Moreover, current models typically rely in many crucial aspects on heuristic assumptions  \citep{Nan2018,Zheng2019} and almost all of them assume simple random fiber networks (e.g., based on Voronoi tesselations) that do not match the specific microstructural characteristics of actual collagen gels or tissues. What remains wanting is a robust, computationally efficient three-dimensional model of cell-fiber interactions, where the microstructure of the fiber network realistically resembles real collagen gels and tissues and which is efficient enough to enable simulations with several cells. Such a computational model can be expected to help unravel the micromechanical and molecular foundations of mechanical homeostasis.

In this paper, we introduce such a computational model. It is based on the finite element method and relies on a strong experimental foundation. It can be used to test various hypotheses with regard to the micromechanical principals of mechanical homeostasis. It can also help to identify promising future experiments. The model focuses on mechanical aspects of homeostasis by concentrating on the physical interactions of cells with surrounding matrix fibers and thus neglects direct modeling of biochemical phenomena. The paper focuses on a detailed description of the computational framework, but examples are used to demonstrate the physical validity of this framework and to illustrate the opportunities it will open up. It will be seen that this framework captures well key observations from experiments on short time scales (in which deposition and degradation of tissue fibers can be neglected) thus helping to explain the underlying physics.

\section{Models and methods}
\label{sec:mat_and_meth}

To study the physical foundations of mechanical homeostasis in soft biological tissues on short time scales (hours), our framework models i) interlinked ECM-like fiber networks whose microstructure closely resembles that of actual collagen gels, ii) transmembrane proteins such as integrins which connect extra- to intra-cellular structures, and iii) the contractile activity of the cytoskeleton. In the following we describe the mathematical and computational details of our model.

\textbf{\subsection{Construction of representative volume elements (RVE)}}
\label{sec:discrete_fiber_model}

\begin{figure*}
\centering
\includegraphics[width=1.0\linewidth]{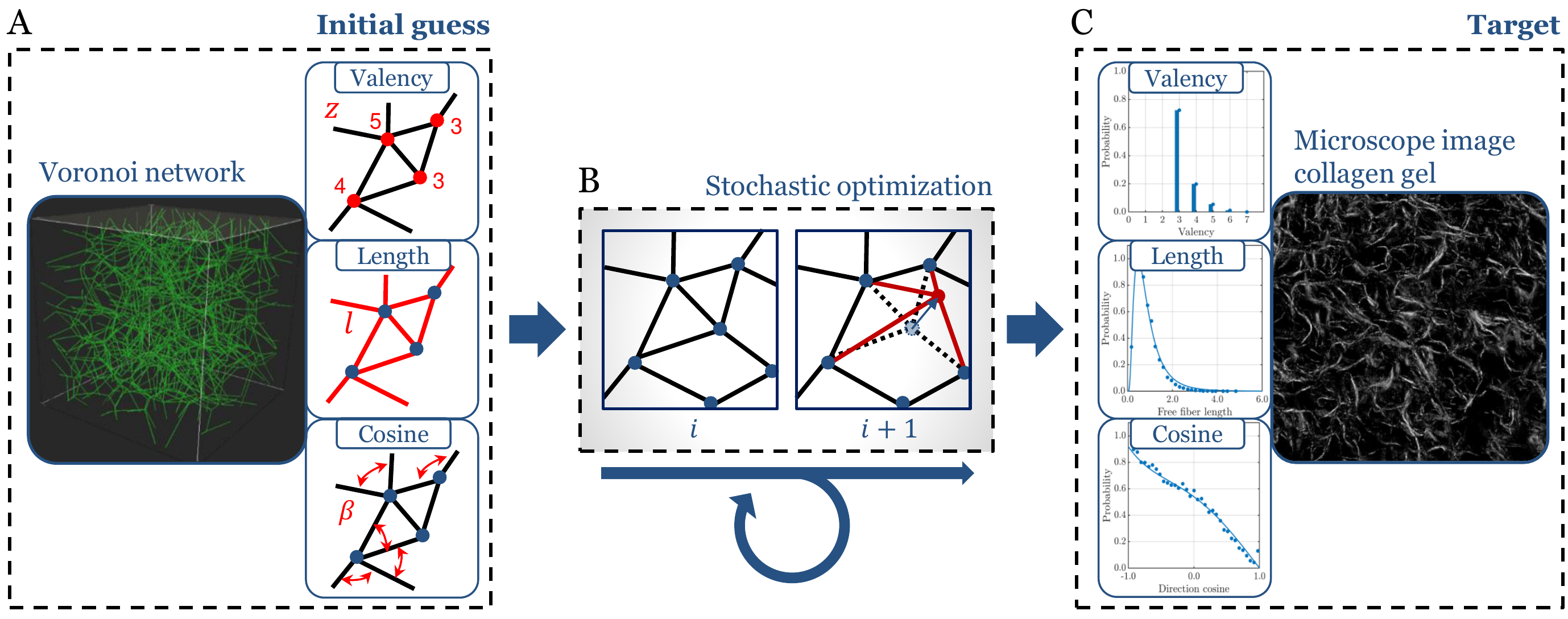}
\caption{Schematic of the network construction process. (A) Random fiber network geometries based on Voronoi tesselation are used as the initial configuration. Valency, length and cosine distribution are used as descriptors of the network geometry for which target distributions are given. (B) By iterative random displacements of arbitrary nodes in the network and accepting these displacements based on their impact on the system energy, which penalizes deviations of the geometric descriptors from their target distributions, one arrives after a number of stochastic steps at a configuration with the desired distribution of the geometric descriptors of interest. (C) Microscope images of collagen gels are used to determine the target distributions for the descriptors of the network.}
\label{fig:network_reconstruction}
\end{figure*}

Computational modeling of soft tissues on the level of discrete fibers and individual cells remains intractable for large tissue volumes, noting that $1\ ml$ of ECM may contain over one million cells. Therefore, we use RVEs as structurally typical samples of the considered tissue (Fig. \ref{fig:network_reconstruction} A). Building on our previous work on biopolymer networks \citep{Cyron2012,Cyron2013,Cyron2013a,GrillParticle2020}, we constructed physically realistic three-dimensional fiber networks from confocal microscope images of actual collagen gels (Fig. \ref{fig:network_reconstruction} A). Following \cite{Lindstrom2010} and \cite{Davoodi-Kermani2021}, we assumed that the mechanical properties of collagen fiber networks are predominantly governed by three descriptors, namely, the valency (number of fibers connected to a network node), the free-fiber lengths between adjacent nodes (herein also referred to as fiber length), and the angles between the fibers joining at the nodes (which can be quantified by the cosine of the angles between any pair of fibers joining at a node). These descriptors vary in the network across the fibers and nodes by following certain statistical distributions. Using the computational procedure described in Appendix \ref{sec:appendix_simanneal}, which is motivated by \cite{Lindstrom2010,Yeong1998} and briefly illustrated in Fig.  \ref{fig:network_reconstruction}, we ensured that the statistical distributions of valency, free-fiber length and inter-fiber cosines closely matched those of actual collagen fiber networks. The computational procedure to produce such networks has been implemented in a short C++ program which is available under the BSD 3-Clause License as the repository \url{bionetgen} hosted at \url{https://github.com/bionetgen/bionetgen}.

\textbf{\subsection{Mechanical network model}}
\label{sec:mechanical_model}

We used the finite element method to model the mechanics of our fibrous RVE. Individual fibers were modeled as geometrically exact beam finite elements based on the non-linear Simo-Reissner theory \citep{Reissner1981,Simo1985,Simo1986} and a hyperelastic material law. This beam theory captures the modes of axial tension, torsion, bending and shear deformation and is appropriate for large deformations. Thus, our finite element model of the fiber network can capture all essential modes of mechanical deformation. If not stated otherwise, covalent bonds between fibers were modeled as rigid joints coupling both translations and rotations. We chose the dimensional and constitutive parameters to mimic collagen type I fibers as the most abundant structural protein of the ECM. Fibers are assumed to have circular cross-sections with a diameter of $D_f = 180\ nm$ \citep{VanDerRijt2006a} and elastic moduli of $E_f = 1.1\ MPa$ \citep{Jansen2018}. Assuming curvilinear fibers with circular cross-section of diameter $D_f $, the average mass density of collagen $\rho_{c}$ in the network RVE was calculated as
\begin{align}
\label{eqn:calc_conc} 
\rho_{c} &= \frac{L_{tot}{D_f}^{2} \pi}{V_{4\text{RVE}} v_{c}} 
\end{align}
according to \cite{Stein2008}, where $ L_{tot} $ is the sum of all individual fiber lengths, $V_{\text{RVE}}$ the volume of the RVE, and $ v_{c} = 0.73 \ \text{ml/g} $ the specific volume of collagen fibers \citep{Hulmes1979}.\\

\textbf{\subsection{Fiber-to-fiber cross-linking}}

A native ECM consists of myriad structural constituents, including collagen and elastin, which usually form networks to provide mechanical support to the resident cells. To form these networks, covalent cross-links are formed via the action of enzymes such as lysol oxidase and transglutaminase, which can be produced by the cells \citep{Simon2014}. In addition to covalent bonds, transient hydrogen bonds or van-der-Waals bonds contribute further to the mechanical integrity of the ECM \citep{Kim2017,Ban2018a}.

\begin{figure*}
\centering
\includegraphics[width=0.8\linewidth]{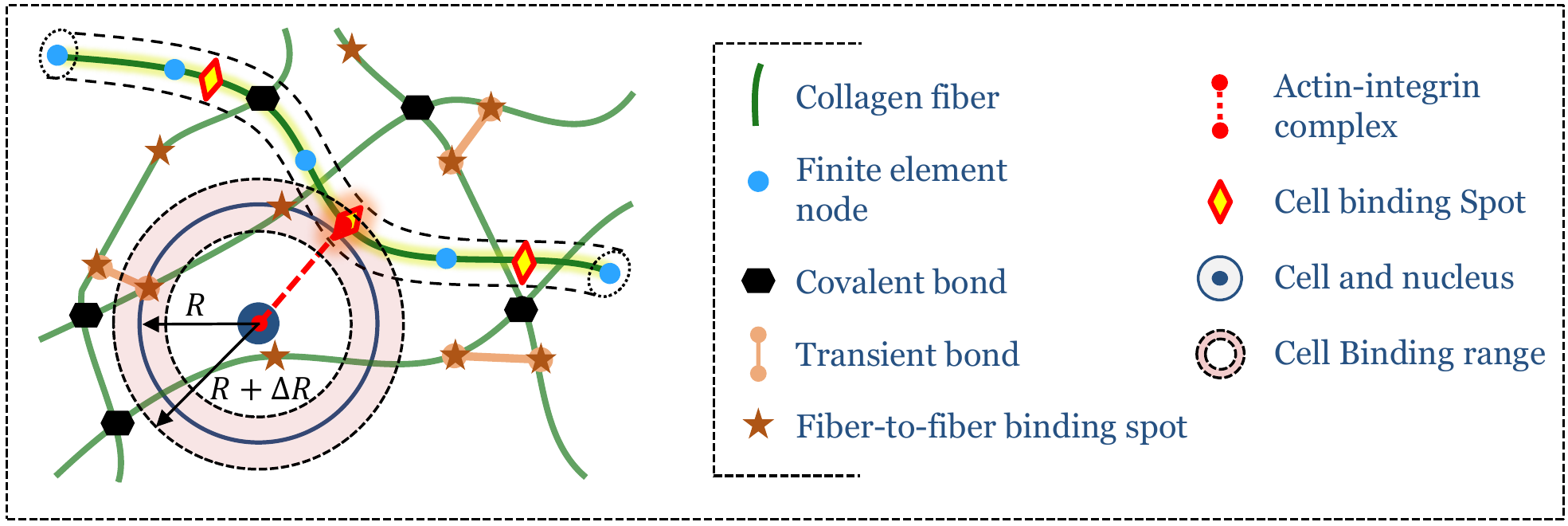}
\caption{Fiber network model: collagen fibers are modeled as beam-like mechanical continua discretized by beam finite elements. Nearby collagen fibers are connected by permanent (covalent) chemical bonds modeled as rigid joints. During the simulation additional transient bonds may stochastically form and dissolve between nearby binding spots of the fibers. These bonds are also modeled by short beam-elements transmitting forces and moments. Cells of radius $R$ can attach to nearby collagen fibers if certain predefined cell binding spots on the surrounding fibers are within $R-\Delta R$ and $R+\Delta R$ around the cell.}
\label{fig:methods}   
\end{figure*}

To model initially existing covalent bonds between fibers, we permanently connect individual fibers joining at nodes of our initially generated network by rigid joints. To model the formation of additional transient and covalent bonds, we define so called binding spots on all fibers (Fig. \ref{fig:methods}). If during the simulation it happens that the distance between two binding spots on distinct filaments falls within a certain critical interval, a new bond between the two filaments is established according to a Poisson process with an on-rate $k_{on}^{f-f}$. That is, within a subsequent time step $\Delta t$, a bond is assumed to form with a probability
\begin{align}
p_{on}^{f-f}&=1 - \exp{(-k_{on}^{f-f}\Delta t)}.
\label{eqn:pon}
\end{align}
Newly established bonds are modeled by initially stress-free beam elements. Bonds established this way during the simulation can also dissolve. This process is again modeled by a Poisson process with an off-rate $k_{off}^{f-f}$, yielding in each time step $\Delta t$ an unbinding probability 
\begin{align}
p_{off}^{f-f}(F)&=1 - \exp{(-k_{off}^{f-f}(F)\cdot \Delta t)}.  
\label{eqn:poff}
\end{align}
The off-rate is in general affected by the force $F$ acting on the bond because transient chemical bonds under mechanical loading are typically less (though in certain regimes more) stable than load-free bonds \citep{Bell1980}. This phenomenon can be modeled by a force-dependent off-rate 
\begin{align}
k_{off}^{f-f}(F) &= k_{off,0}^{f-f}\  \exp{\left( \frac{F \Delta x}{k_B T}\right) }, 
\label{eqn:koff_f}
\end{align}
with $\Delta x$ a characteristic distance, $k_B$ the Boltzmann constant, and $T$ the absolute temperature \citep{Bell1980}. $\Delta x > 0$ was chosen so that the bond weakens under tension, a bond behavior that is often referred to as slip-bond behavior. By choosing $k^{f-f}_{off,0} = 0$, we can resemble new covalent bonds formed during our simulations, whereas $k^{f-f}_{off,0} > 0$ mimics transient bonds. 

\textbf{\subsection{Cell-ECM interaction}}
\label{subsec:cell_ecm}

\begin{figure*}
\centering
\includegraphics[width=0.80\linewidth]{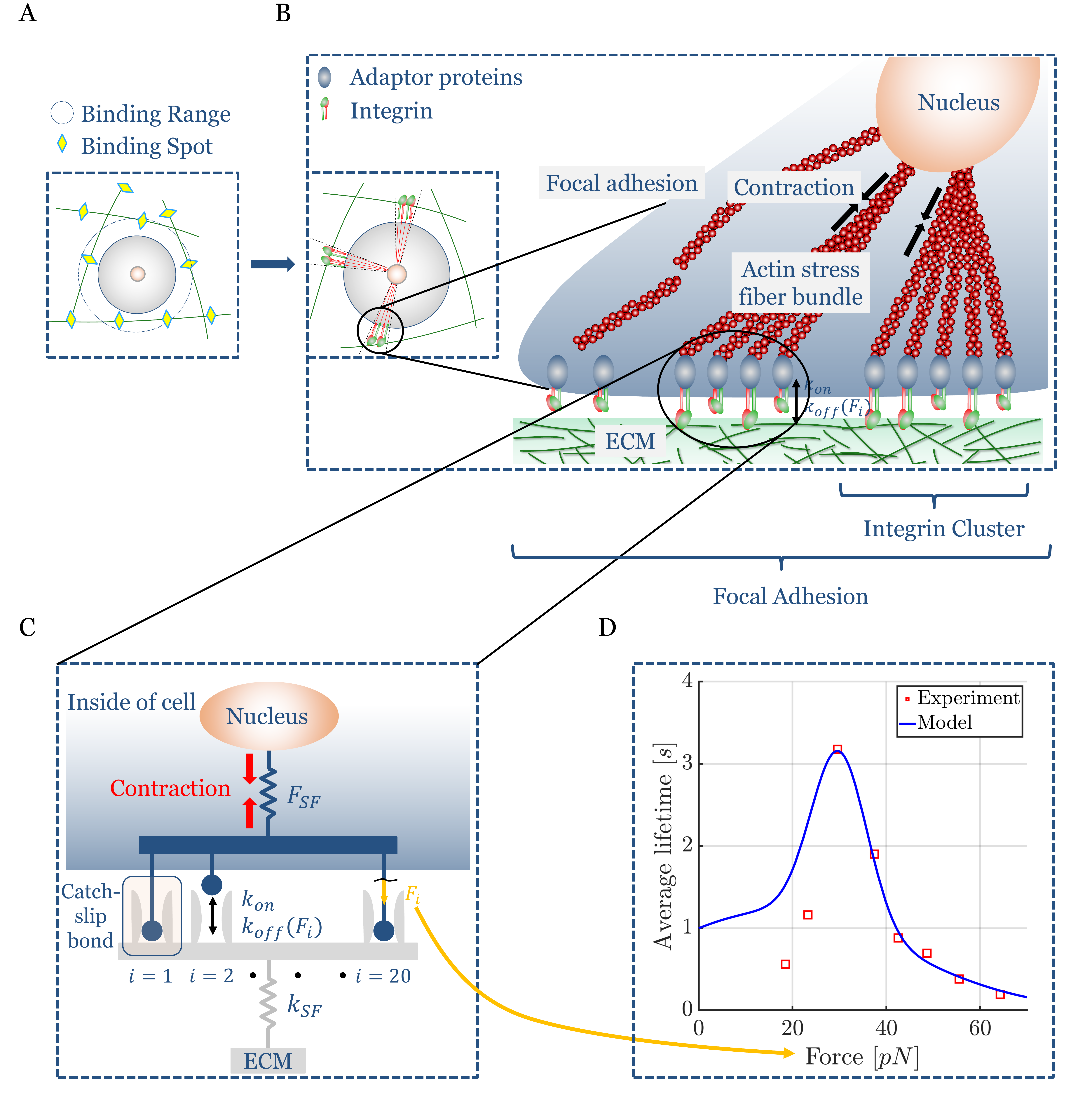}
\caption{(A) If cells lie within a certain distance from integrin binding spots on fibers, a focal adhesion can from with a certain probability. (B) A focal adhesion consists of around 1000 integrins connecting the intra-cellular actin cytoskeleton to the ECM fibers. Actin stress fibers connect the cell nucleus to the focal adhesions and are modeled as elastic springs that contract over time. (C) Each focal adhesion consists of numerous so-called integrin clusters, each formed by $20-50$ integrins. We assume that each integrin cluster is connected to one actin stress fiber. Integrins are modeled as molecular clutches, i.e., they bind and unbind according to specific binding kinetics. (D) Experiments have determined a catch-slip bond behavior for single integrins where the life time does not monotonically decrease with the mechanical force transmitted through the bonds but where there exists a regime where increasing forces increase the average life time of the bond. To avoid infinite off-rates in case of low forces, we chose a slightly higher life time for low forces compared to the experimental data of \cite{Kong2009}.}
\label{fig:cell_ecm}
\end{figure*}

Cells in soft tissues can mechanically connect to surrounding fibers by integrins and exert stress on them by focal adhesions. A focal adhesion usually includes an actin stress fiber bundle in the cytoskeleton that connects the nucleus of the cell with the integrins of a cluster and can actively contract. Based on experimental observations, we restricted the maximal number of focal adhesions per cell to $N_{FA,max}=65$ \citep{Horzum2014,Kim2013,Mason2019} (Fig. \ref{fig:cell_ecm} B on the left shows 3 focal adhesions). It has been shown experimentally that roughly $N_{i,FA,max} = 1000$ integrins are involved in one focal adhesion \citep{Wiseman2004,Elosegui-Artola2014}. These integrins are organized in so-called integrin clusters of roughly $ 20-50$ integrins \citep{Changede2015,Cheng2020} (Fig. \ref{fig:cell_ecm} C). We thus assume for each focal adhesion 50 integrin clusters containing a maximum of $N_{i,ic,max} = 20$ integrins each.

To model cell-mediated active mechanical processes in soft tissues, we model the cell centers as point-like particles. When these particles approach predefined integrin binding spots (with a distance of $d^{i-f} = 50 nm$ to each other \citep{Lopez-Garcia2010}) on the fibers within $\pm \Delta R$ around the cell radius $R$, a physical connection between cells and fibers is assumed to form by a Poisson process similar to the one in Eq. \eqref{eqn:pon}, but with a specific on-rate $k_{on}^{c-f}$ (see also Fig. \ref{fig:cell_ecm} A). The actin stress fibers connecting the cell nucleus with the fibers surrounding the cells are modeled as elastic springs (Fig. \ref{fig:cell_ecm} B and C) whose stress-free length evolves at some predefined rate $\dot{c}$ that can be calculated to match experimental data of different cell types. These stress fiber contract at a rate of  $\dot{c} = 0.1 \frac{\mu m}{s}$ \citep{Choquet1997,Moore2010}. The force acting on a single integrin $F_i$ can be computed according to 
\begin{align}
F_i = \frac{F_{SF}}{N_{i,bonded}},
\label{eqn:force_acting_on_integrin}
\end{align}
with $F_{SF}$ being the force acting in the respective stress fiber and $N_{i,bonded}$ the number of currently bound integrins in the integrin cluster associated with the respective stress fiber.

In contrast to many previous approaches in which displacements have been prescribed in the neighborhood of cells to model their contraction, we are able to model a true two-way feedback loop between cell and ECM. Integrins have been shown experimentally to exhibit a so-called catch-slip bond behavior \citep{Kong2009} whose unbinding can be modeled by a Poisson process with a force-dependent off-rate
\begin{equation}
\begin{split}
k_{off}^{c-f}(F) = a_1\text{exp}\left( -\left( \frac{F-b_1}{c_1}\right) ^2\right)  \\ + a_2\text{exp}\left( -\left( \frac{F-b_2}{c_2}\right) ^2\right)
\end{split}
\label{eqn:integrin_koff}
\end{equation}
whose parameters were determined via fits to the experimental data \citep{Kong2009,Weng2016} (Fig. \ref{fig:cell_ecm} D) and can be found in Table \ref{table:model_parameters}. While the average lifetime of most chemical bonds decreases monotonically with increasing force transmitted by the bond, catch-slip bonds exhibit a regime where the bond stabilizes as the force increases. As illustrated in Fig. \ref{fig:cell_ecm} D, this makes integrin bonds particularly stable for values of $F_i$ in a range around $30 pN$. Recall that we model an integrin cluster as a system of 20 parallel integrins whose bonds form and dissolve according to the above specified on- and off-rates (Fig. \ref{fig:cell_ecm} C). If at a certain point all bonds happen to have broken at the same time, the related integrin cluster is assumed to dissolve. It may, however, reform on the basis of a new (not yet contracted) stress fiber shortly thereafter with a binding rate $k_{on}^{c-f}$. If all clusters of a certain focal adhesion happen to dissolve at the same time, the focal adhesion as a whole is dissolved. 

This model implies that many binding and unbinding events of integrins occur during the lifetime of a focal adhesion. This way, our model captures the chemical dynamics of the connection between cells and ECM fibers on different scales ranging from individual integrins to whole focal adhesions \citep{Stehbens2014}. Thereby, our model bots captures typical lifetimes of focal adhesions on the order of minutes and turnover rates of most proteins involved in the adhesion complex on the order of seconds. 

\textbf{\subsection{Boundary conditions}}
\label{subsec:boundary_cond}

As mentioned before, simulations of complete tissues on the $cm$-scale are computationally expensive with discrete fiber models; hence we have to study RVEs. A major challenge in the context of discrete fiber simulations is the imposition of deformations on the RVE to study its response to certain strains. To this end, most previous work of other groups requires that the nodes of the finite elements used to discretize fibers are located exactly on the boundary surfaces of the RVE where displacements are prescribed \citep{Stein2011,Humphries2018,Abhilash2014,Burkel2018,Ban2018a,Liang2016a,Ban2019}. Other approaches prescribe the displacements of nodes close to these surfaces \citep{Lee2014}. These methods share the problem that they do not ensure full periodicity across the boundaries where displacements are prescribed. To overcome this limitation, we developed a novel form of fully periodic boundary conditions for fiber networks that allows the imposition of complex multiaxial loading states. This approach ensures full periodicity across all surfaces of the RVE and thereby minimizes computational artifacts due to finite-volume effects. The computational details of our algorithm are summarized in Appendix \ref{sec:boundary_conditions}.

\textbf{\subsection{Search algorithm and parallel computing}}
\label{subsec:parallel_computing}

To yield meaningful computational results, our RVEs have to be much larger than the characteristic microstructural features such as the free-fiber length between adjacent nodes. Using values for the cell density and collagen concentration in a physiologically reasonable range typically leads to a system size of the RVE that can be solved only by parallel computing, including an efficient parallel search algorithm for the evaluation of all interactions between cells and fibers. We implemented such a search algorithm based on a geometrical decomposition of the computational domain in uniform cubic sub-domains. The computational details of our parallelization are summarized in Appendix \ref{sec:par_search}. Importantly, our approach does not require any fully redundant information on all processes, which enables a highly efficient parallelization on even a very large number of processors.\\

\section{Results and discussion}
\label{sec:results}

The presented computational framework was implemen\-ted in our in-house research finite element code BACI \citep{Baci2020}. To ensure robustness, scalability and especially validity, we performed various computational simulations and compared the results with available experimental data. The default parameters used in our simulations can be found in Table \ref{table:model_parameters}. 

\textbf{\subsection{Network construction}}

\begin{figure*}[htbp] 
\begin{minipage}[b]{0.0\linewidth}
\end{minipage}
\hfill
\begin{minipage}[b]{0.0\linewidth}
\end{minipage}
\hfill
\begin{minipage}[b]{0.32\linewidth}
A
\end{minipage}
\hfill
\begin{minipage}[b]{0.32\linewidth}
B
\end{minipage}
\hfill
\begin{minipage}[b]{0.32\linewidth}
C
\end{minipage}
\hfill
\begin{minipage}[b]{0.0\linewidth}
\end{minipage}
\hfill
\\
\begin{minipage}[b]{0.0\linewidth}
\end{minipage}
\hfill
\begin{minipage}[b]{0.32\linewidth}
	\centering
	\includegraphics[width=1.0\linewidth]{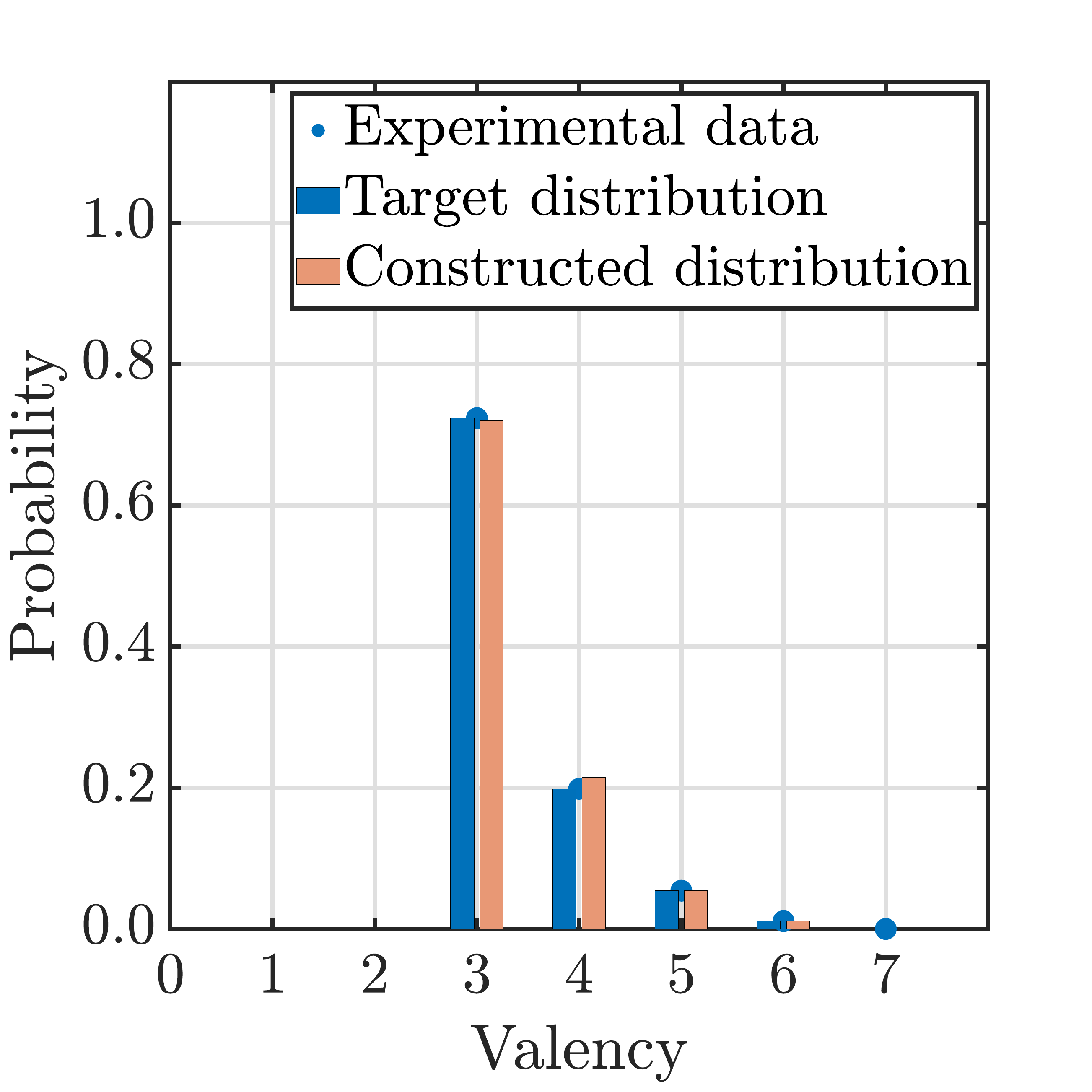}
\end{minipage}
\hfill
\begin{minipage}[b]{0.32\linewidth}
	\centering
	\includegraphics[width=1.0\linewidth]{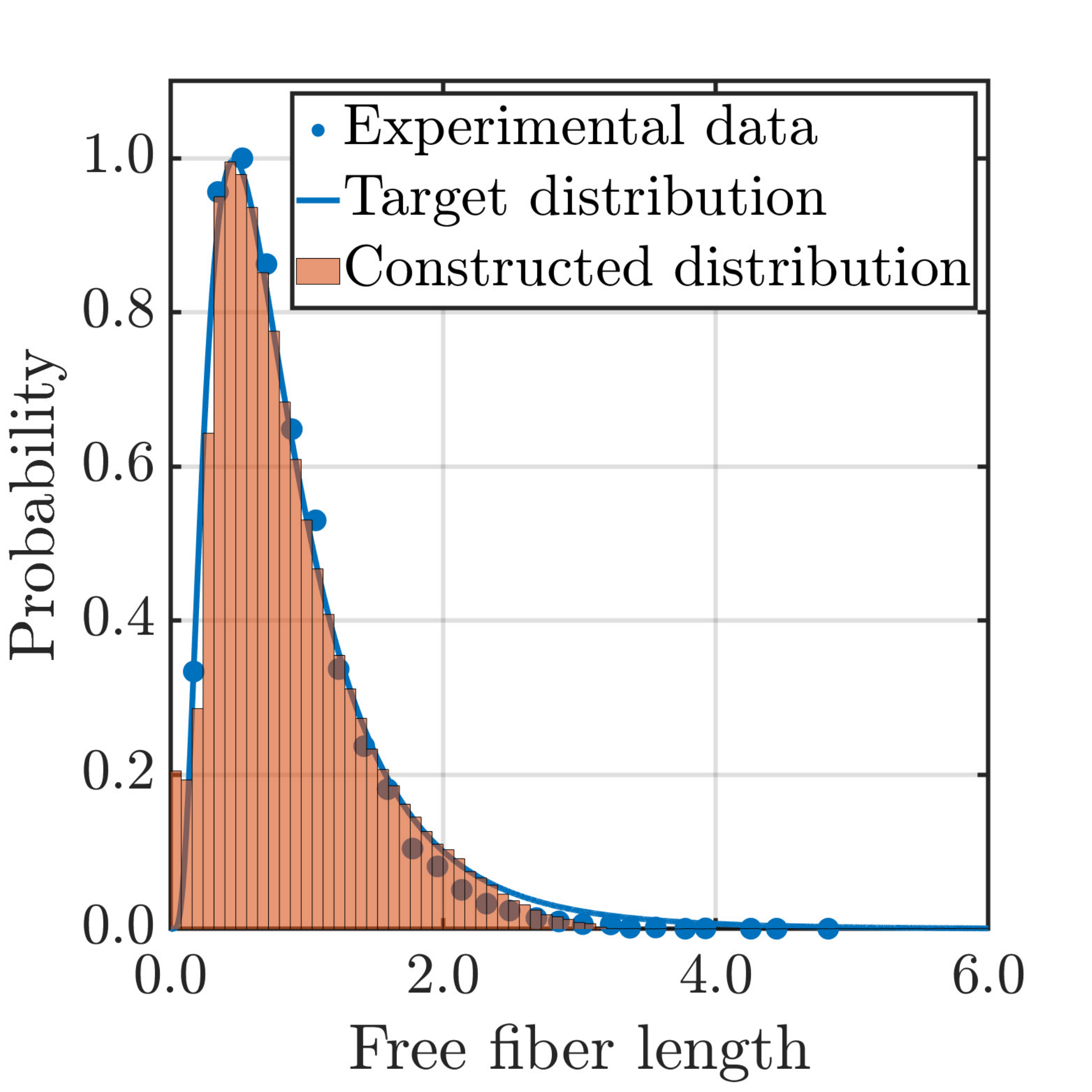}
\end{minipage}
\hfill
\begin{minipage}[b]{0.32\linewidth}
	\centering
	\includegraphics[width=1.0\linewidth]{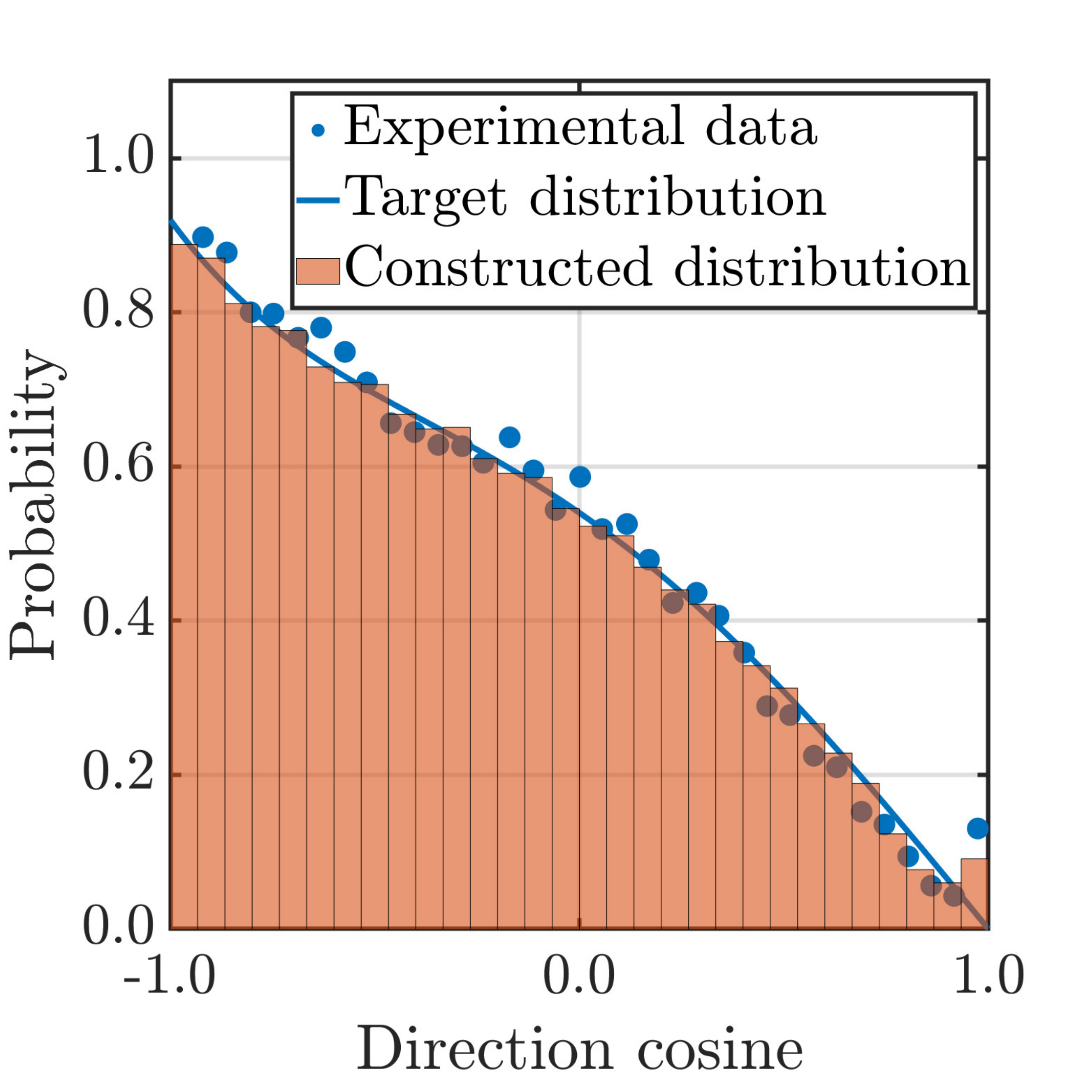}
\end{minipage}
\hfill
\begin{minipage}[b]{0.0\linewidth}
\end{minipage}
\hfill
\caption{Results of the network construction process for a collagen concentration of $2.5 \textit{mg/ml}$. (A) valency distribution,  (B) free-fiber length distribution and (C) cosine distribution fit well the target distributions defined on the basis of experimental data taken from \cite{Nan2018} in (A) and from \cite{Lindstrom2010} in (B) and (C).}
\label{fig:sim_annealing_results}
\end{figure*}

\begin{figure*}[htbp] 
\begin{minipage}[b]{0.0\linewidth}
\end{minipage}
\hfill
\begin{minipage}[b]{0.0\linewidth}
\end{minipage}
\hfill
\begin{minipage}[b]{0.32\linewidth}
A
\end{minipage}
\hfill
\begin{minipage}[b]{0.32\linewidth}
B
\end{minipage}
\hfill
\begin{minipage}[b]{0.32\linewidth}
C
\end{minipage}
\hfill
\newline
\begin{minipage}[b]{0.0\linewidth}
\end{minipage}
\hfill
\begin{minipage}[b]{0.32\linewidth}
	\centering
	\includegraphics[width=1.0\linewidth]{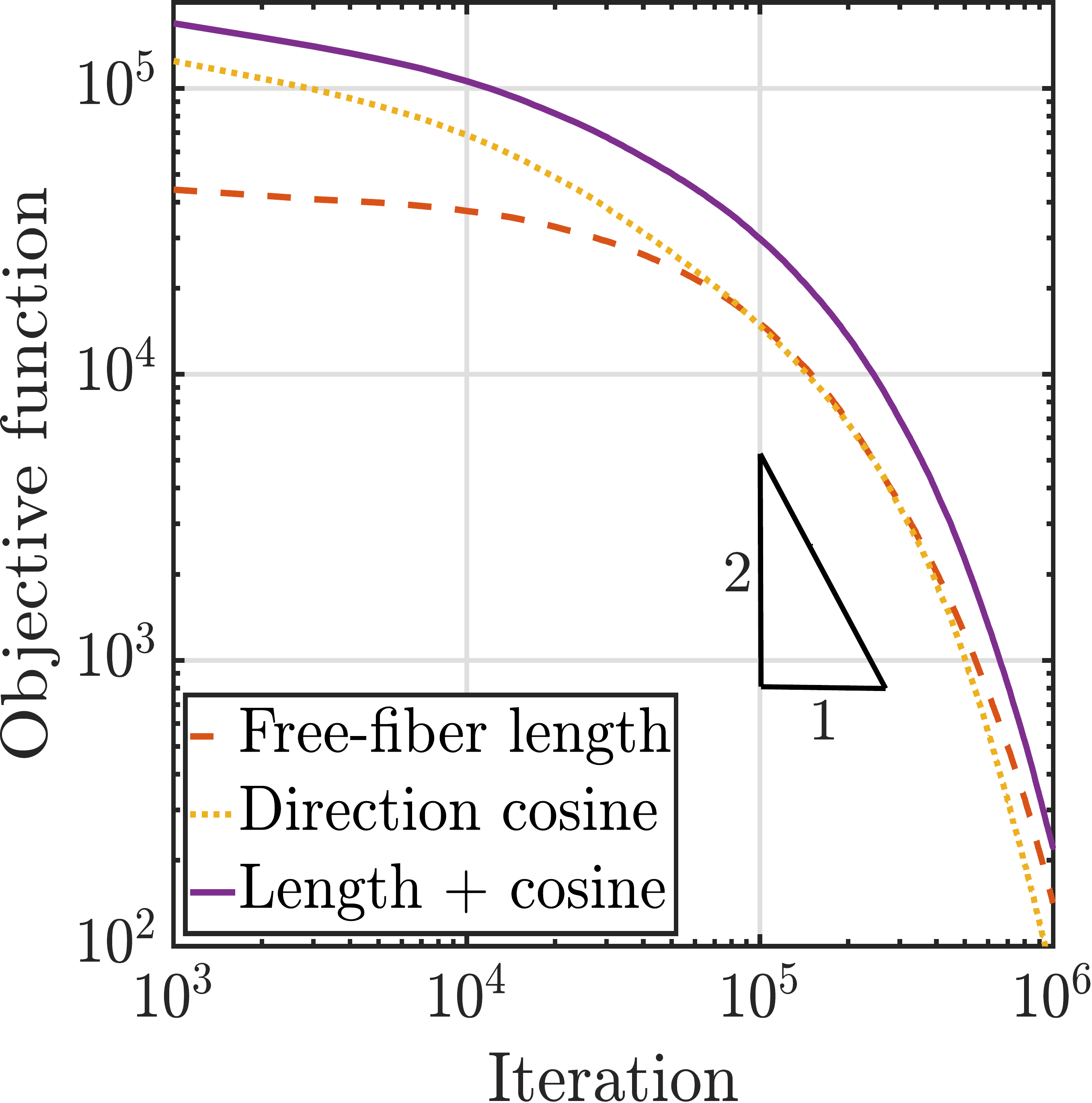}
\end{minipage}
\hfill
\begin{minipage}[b]{0.32\linewidth}
	\centering
	\includegraphics[width=1.0\linewidth]{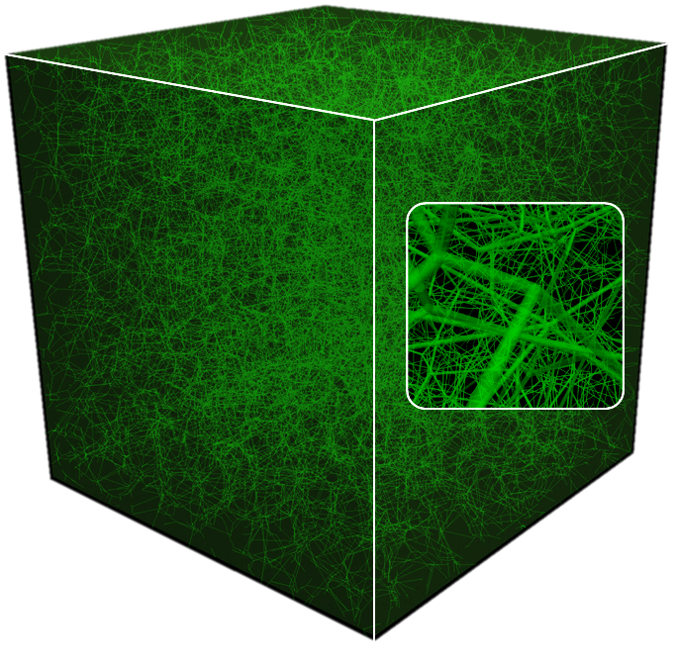}
\end{minipage}
\hfill
\begin{minipage}[b]{0.0\linewidth}
\end{minipage}
\hfill
\begin{minipage}[b]{0.32\linewidth}
	\centering
	\includegraphics[width=1.0\linewidth]{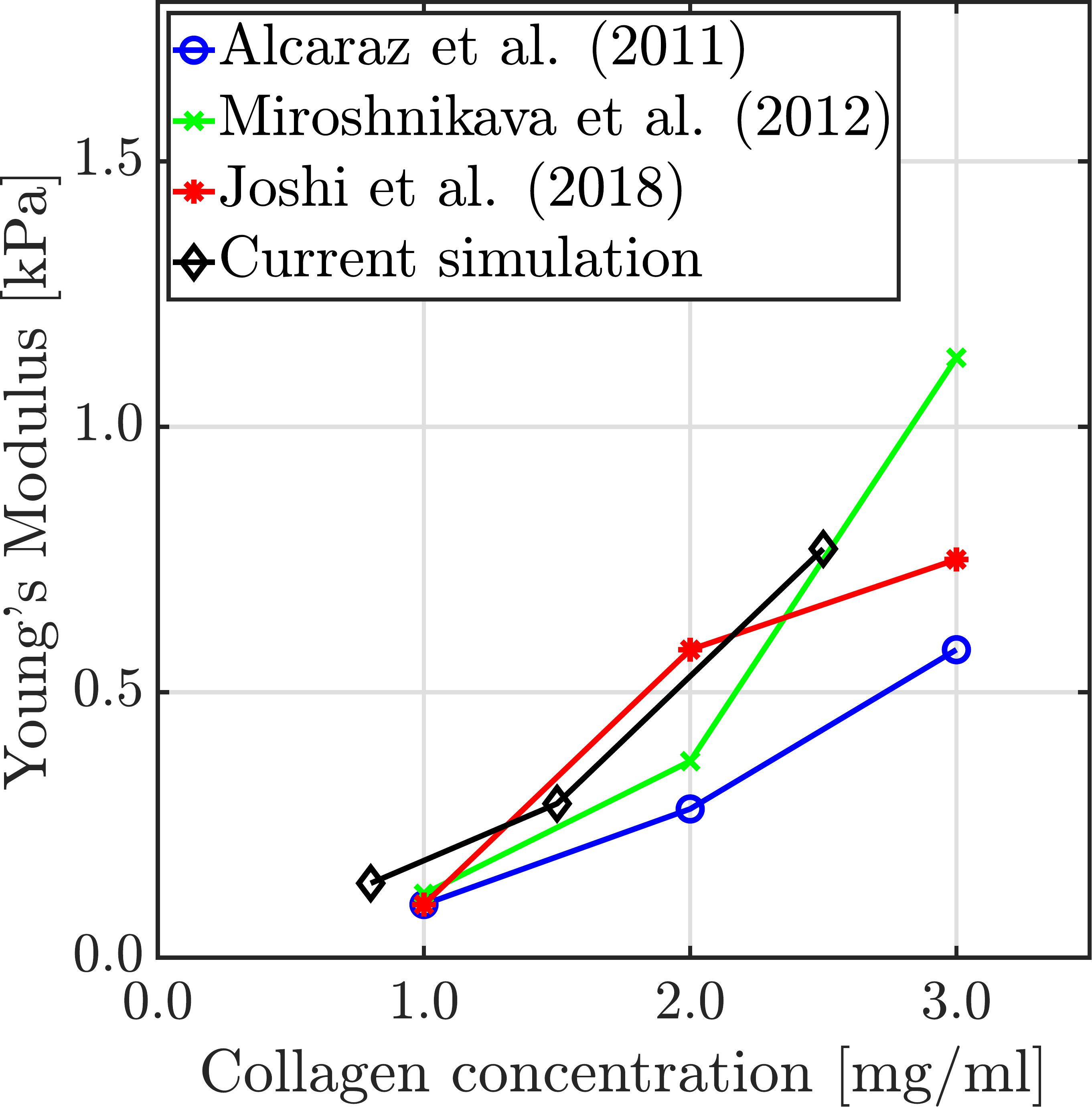}
\end{minipage}
\hfill
\caption{(A) In the stochastic network construction with a collagen concentration of $0.8 \text{mg/ml}$ in a cube of edge length $245 \mu m$, the energy-type objective function according to Eq. \eqref{eqn:energy} is reduced during simulated annealing (in the studied range even superquadratically) by multiple orders of magnitude; (B) this optimization process yields RVE with a desired microstructure; (C) the effective Young's modulus at strains $< 1\%$ of RVE constructed this way match well with the ones observed in experiments \citep{Joshi2018,Alcaraz2011,Miroshnikova2011}.
}
\label{fig:ev_of_energy}
\end{figure*}

We first validated the network generation method described in Section \ref{sec:discrete_fiber_model}. To this end, we created networks with different collagen concentrations and target descriptor distributions as observed by confocal microscopy in tissue culture experiments with collagen type I gels \citep{Lindstrom2010,Nan2018}. As shown in 
Fig. \ref{fig:sim_annealing_results}, our stochastic optimization method successfully generates networks with the desired distributions of valency, free-fiber length and direction cosine. Fig. \ref{fig:ev_of_energy} A demonstrates that our simulated annealing converged well toward the desired solution with an increasing number of random iteration steps. 

\textbf{\subsection{Passive mechanical properties: stiffness}}
\label{subsec:uniaxial_test}

Next, we verified that our constructed, still acellular, networks have similar mechanical properties as actual collagen networks. To this end, we simulated simple uniaxial tensile tests with different collagen concentrations and compared the resulting values for the stiffness with values that have been collected in uniaxial experiments with collagen type I gels \citep{Joshi2018,Alcaraz2011,Miroshnikova2011}. We stretched a cubic simulation box with edge length $L = 245 \mu m$ in one direction by applying displacement boundary conditions as described in Appendix \ref{sec:boundary_conditions} at a slow loading rate of $0.01 \mu \text{m/s} $ up to a strain of $1.0 \%$; strains around $1 \%$ have been shown to be the relevant range when studying active, cell-mediated force development \citep{Eichinger2020}. Fig. \ref{fig:ev_of_energy} C demonstrates that the Young\textquotesingle s moduli of the constructed networks match well with the values observed in tissue culture experiments. In our artificial RVE we found a power law dependence between the Young's modulus and the collagen concentration with an exponent of $1.33$, very similar to the exponent of $1.22$ found experimentally \citep{Joshi2018}.

\textbf{\subsection{Active mechanical properties: homeostatic tension}}

In this section we consider cell-seeded fiber networks to study the \textit{active} mechanics of soft tissues. The tension that develops in constrained gels stems from the contractile forces exerted by the cells on the surrounding fibers. In initially stress-free collagen gels seeded with fibroblasts, the tension builds up over a couple of hours until it has reached a plateau value, the so-called homeostatic value \citep{Marenzana2006,Brown1998,Ezra2010,Eichinger2020,Brown2002,Courderot-Masuyer2017,Campbell2003a,Dahlmann-Noor2007,Karamichos2007,Sethi2002}. Tissue culture experiments \citep{Eichinger2020,Delvoye1991} have shown that the homeostatic tension depends on both cell and collagen concentration in the gel. We used this observation to validate our computational model. We created RVEs with an edge length of $L = 245 \mu m$ and three different cell densities and collagen concentrations as studied experimentally in \cite{Eichinger2020}. To increase the complexity of the RVE only gradually by adding cells, we still solely considered covalent bonds between matrix fibers. We then compared the cell-mediated active tension over time of our simulations to the one observed experimentally.

It is important to note that a direct (quantitative) comparison between experimental data and simulation results is difficult due to differing boundary conditions. Tissue culture experiments have at least one (traction-) free boundary (uniaxial gels have two, circular discs three), while we performed our simulations with RVEs with periodic boundary conditions applied in all directions (note also that a free boundary in a microscopic RVE would not resemble a free boundary of a macroscopic specimen). It has been shown, however, that the number of fixed boundaries has a crucial impact on the homeostatic plateau value \citep{Eichinger2020}. In the following, we compare the first Piola-Kirchhoff stresses, as the thickness of the gel samples over time is unknown. We assumed $A_{initial}=1.6mm$ (knowing it to be between $1.5mm$ and $2mm$) to fit best to our simulation data.

\textbf{\subsubsection{Variation of cell density}}

\begin{figure*}[htbp] 
\begin{minipage}[b]{0.0\linewidth}
\end{minipage}
\hfill
\begin{minipage}[b]{0.32\linewidth}
A
\end{minipage}
\hfill
\begin{minipage}[b]{0.32\linewidth}
B
\end{minipage}
\hfill
\begin{minipage}[b]{0.32\linewidth}
C
\end{minipage}
\hfill
\begin{minipage}[b]{0.0\linewidth}
\end{minipage}
\hfill
\\
\begin{minipage}[b]{0.0\linewidth}
\end{minipage}
\hfill
\begin{minipage}[b]{0.32\linewidth}
	\centering
	\includegraphics[width=1.0\linewidth]{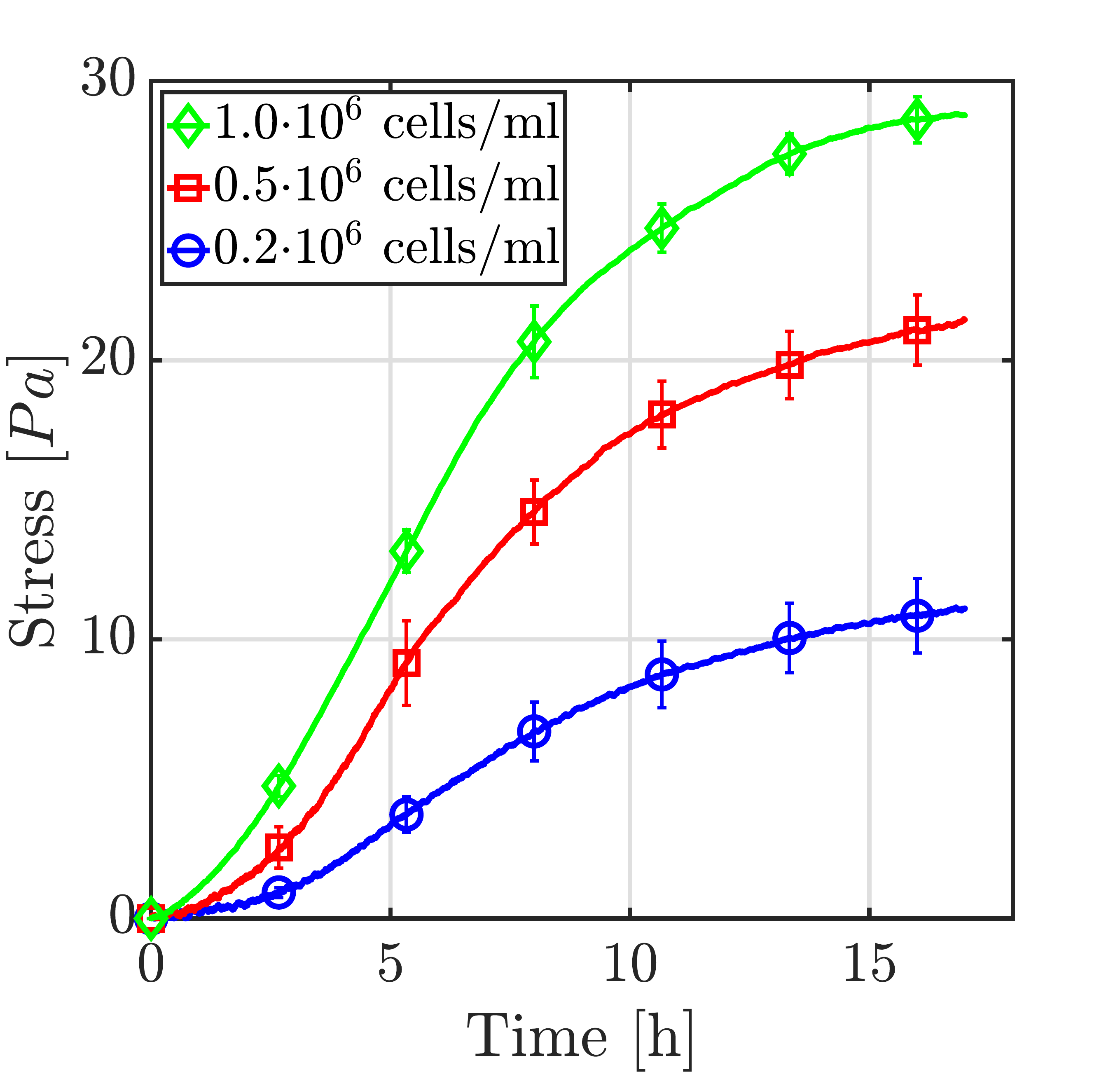}
\end{minipage}
\hfill
\begin{minipage}[b]{0.32\linewidth}
	\centering
	\includegraphics[width=1.0\linewidth]{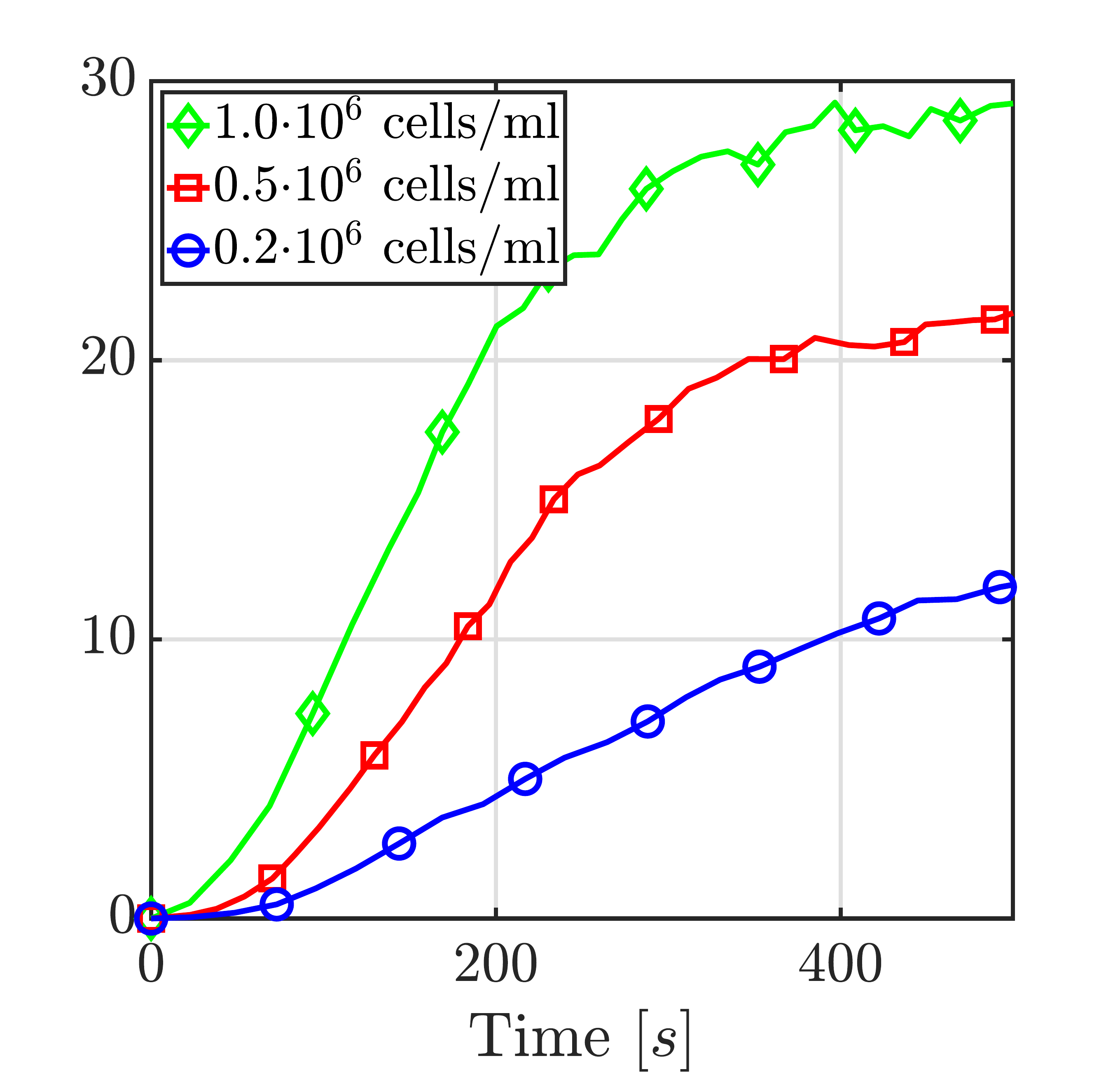}
\end{minipage}
\hfill
\begin{minipage}[b]{0.32\linewidth}
	\centering
	\includegraphics[width=1.0\linewidth]{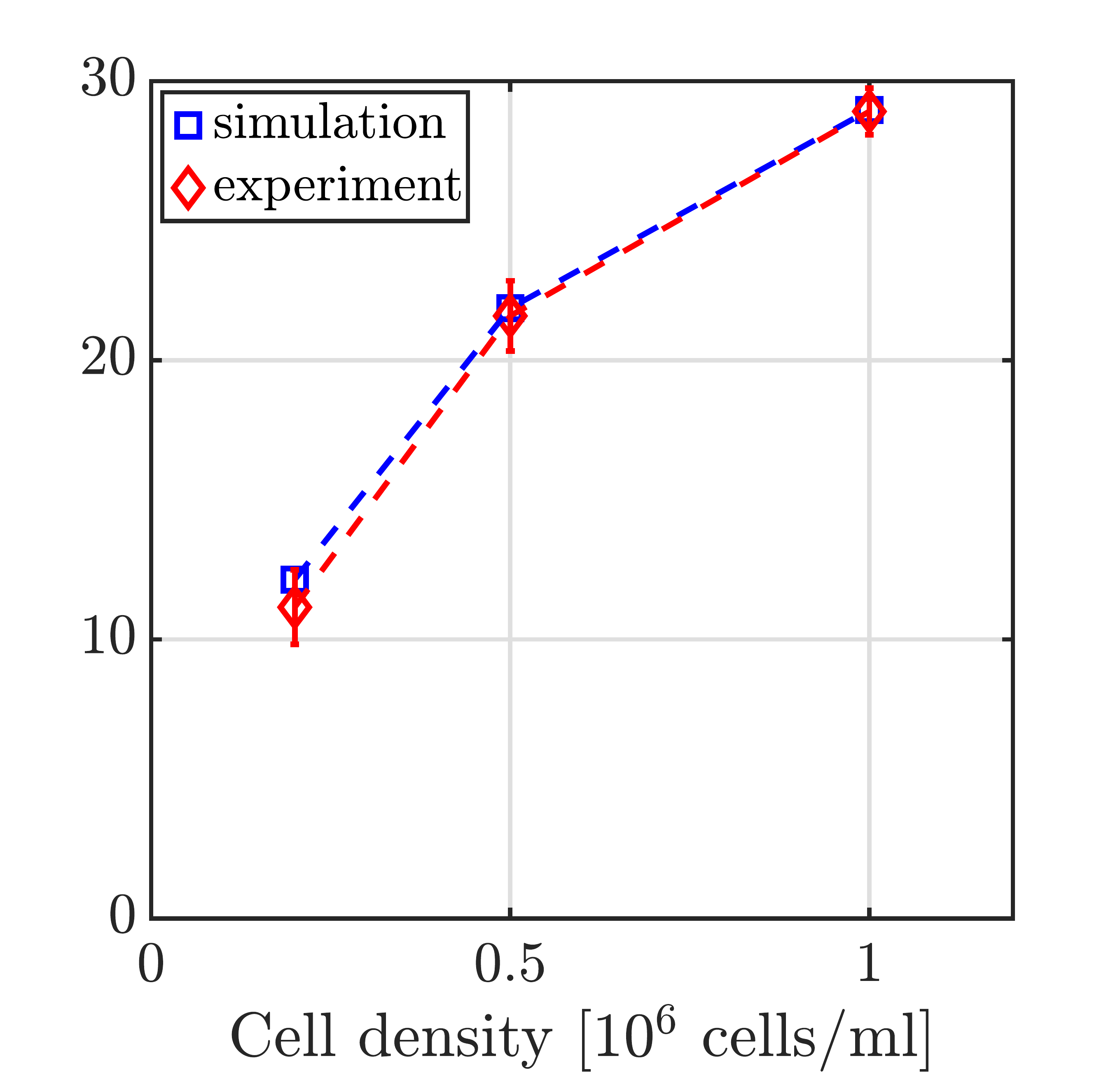}
\end{minipage}
\hfill
\begin{minipage}[b]{0.0\linewidth}
\end{minipage}
\hfill
\caption{For a collagen concentration of $1.5 \text{mg/ml}$, we compare the development of the first Piola-Kirchhoff stress in (A) experiments \citep{Eichinger2020} and (B) simulations. A good semi-quantitative agreement of the expected cell-mediated steady state with non-zero tension (last data points of (A) and (B)) is observed (C), however, also a significant difference of the time scales.}
\label{fig:cell_density}
\end{figure*}

In this section we consider gels with a constant collagen density of $1.5 mg/ml$. Cell densities of $0.2 \cdot 10^6\ \text{cells/ml}$, $0.5 \cdot 10^6\ \text{cells/ml}$ and 1.0 $\cdot 10^6\ \text{cells/ml}$ studied in \cite{Eichinger2020} translate in our simulations into 3, 8 and 15 cells per RVE, respectively. Fig. \ref{fig:cell_density} A shows the evolution of the first Piola-Kirchhoff stress (true force/original area) generated in uniaxially constrained, dog-bone sha\-ped collagen gels as observed experimentally. The gradient during the first $10h$ of the experiment and the homeostatic plateau level of stress increase with the cell density. Both features are observed also in our simulations and fit quantitatively well (Fig. \ref{fig:cell_density} B and C). We can therefore conclude that actin cytoskeleton contraction along with the focal adhesion dynamics described in Section \ref{subsec:cell_ecm}.4 are sufficient mechanisms to reproduce this non-trivial relationship.

A crucial difference between experiments and simulations is the time scale. Whereas mechanical homeostasis develops over a couple of hours in the experiments, it does so within a couple of minutes in the simulation. Interestingly, this time scale of our simulations agrees well with that for which single cells in experiments on purely elastic substrates reach a homeostatic state \citep{Weng2016,Hippler2020}. Thus, a possible explanation for the difference between our simulations and the experimental data from \cite{Eichinger2020} may be that in tissues with numerous cells, complex interactions between the cells substantially delay the homeostatic state. Such interactions remain poorly understood and are not yet accounted for in our computational framework. Another possible explanation for the different time scales in Fig. \ref{fig:cell_density} A and B may be viscoelasticity due to collagen fibers moving within culture media, which is not included in our model in detail, and due to an increasing stiffness of the gel due to progressed polymerization when being placed in an incubator of 37\textdegree C for longer times. Finally, subtle aspects on the sub-cellular scale that are not included in our model may affect the time to reach the homeostatic state substantially because it is well-known that this time differs considerably for different cell types \citep{Eichinger2020b}. 

\begin{figure*}
\centering
\includegraphics[width=0.95\linewidth]{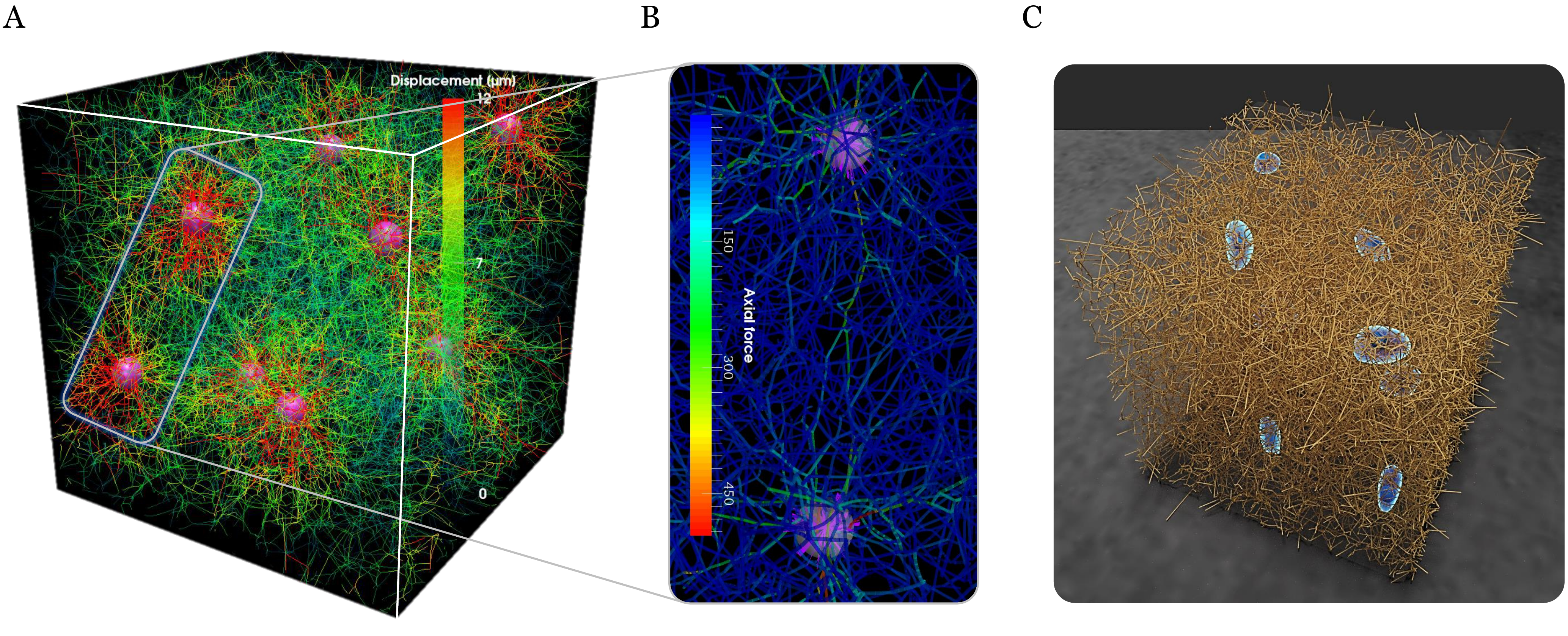}
\caption{Cells mechanically interact with surrounding matrix fibers. (A) Cells attach to nearby fibers, contract and thereby deform the matrix. The simulated, cell-mediated matrix displacements are in a realistic range when compared to experimental data \citep{Notbohm2015,Malandrino2019}. (B) Contracting cells can mechanically interact with other cells over a distance of several cell diameters via long-range mechanical signaling through matrix fibers, a phenomenon observed also in experiments \citep{Kim2017,Ma2013,Shi2014,Baker2015,Mann2019}. (C) Cells, visualized with reconstructed cell membrane around stress fibers, develop different shapes when pulling on the ECM.}
\label{fig:paraview_cells_in_networks}
\end{figure*}

Fig. \ref{fig:paraview_cells_in_networks} A shows that the deformation of the matrix fibers around the cells in our simulations are on the order of $10 \mu m$, which agrees well with experiments
\citep{Notbohm2015,Malandrino2019}. Our simulation framework also reproduces the ability of cells to communicate via long-range mechanical interactions over several cell diameters (Fig. \ref{fig:paraview_cells_in_networks} B), which has also been observed experimentally \citep{Kim2017,Ma2013,Shi2014,Baker2015,Mann2019}. 

\textbf{\subsubsection{Variation of collagen concentration}}

\begin{figure*}[htbp] 
\begin{minipage}[b]{0.0\linewidth}
\end{minipage}
\hfill
\begin{minipage}[b]{0.32\linewidth}
A
\end{minipage}
\hfill
\begin{minipage}[b]{0.32\linewidth}
B
\end{minipage}
\hfill
\begin{minipage}[b]{0.32\linewidth}
C
\end{minipage}
\hfill
\begin{minipage}[b]{0.0\linewidth}
\end{minipage}
\hfill
\\
\begin{minipage}[b]{0.0\linewidth}
\end{minipage}
\hfill
\begin{minipage}[b]{0.32\linewidth}
	\centering
	\includegraphics[width=1.0\linewidth]{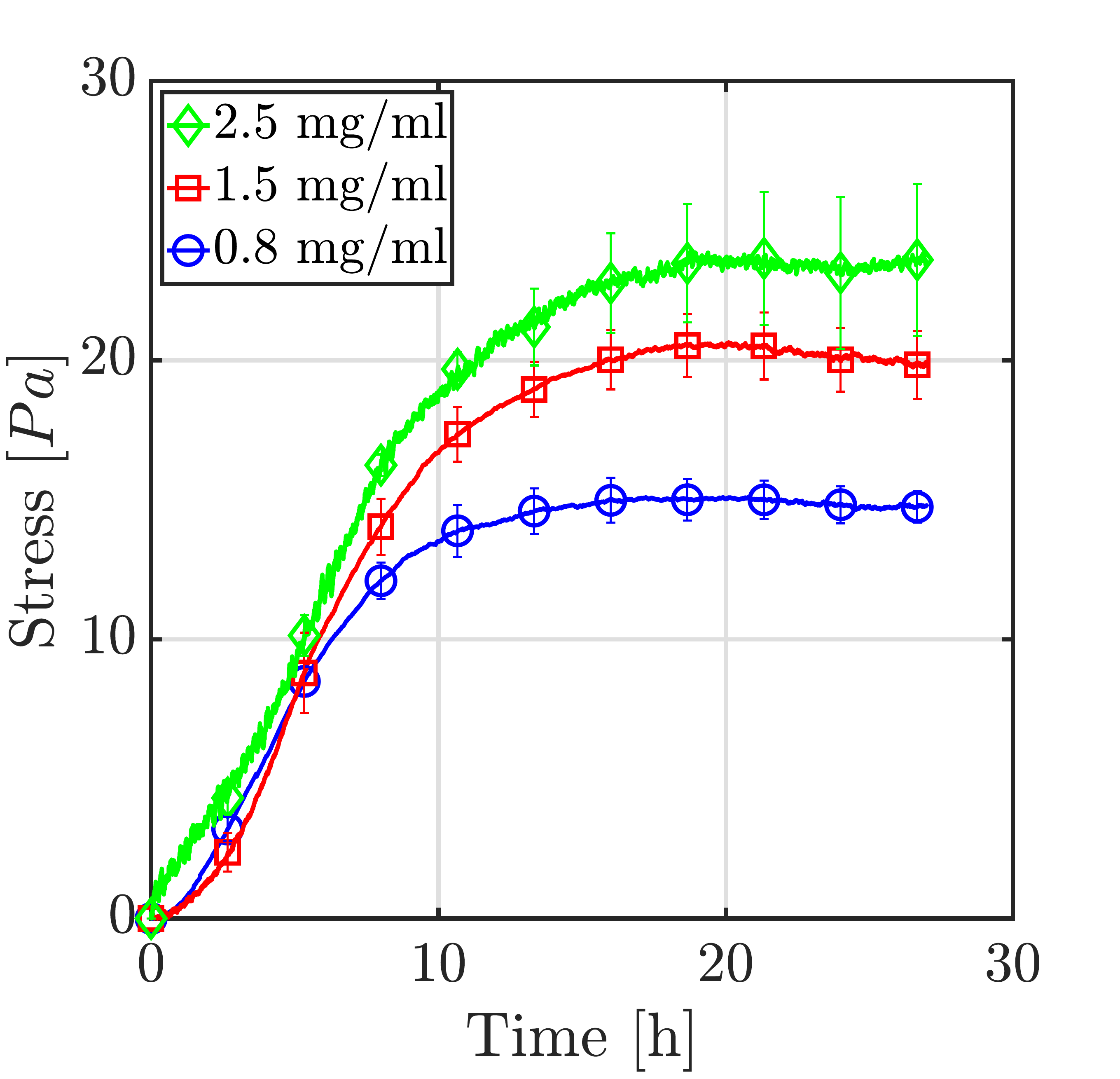}
\end{minipage}
\hfill
\begin{minipage}[b]{0.32\linewidth}
	\centering
	\includegraphics[width=1.0\linewidth]{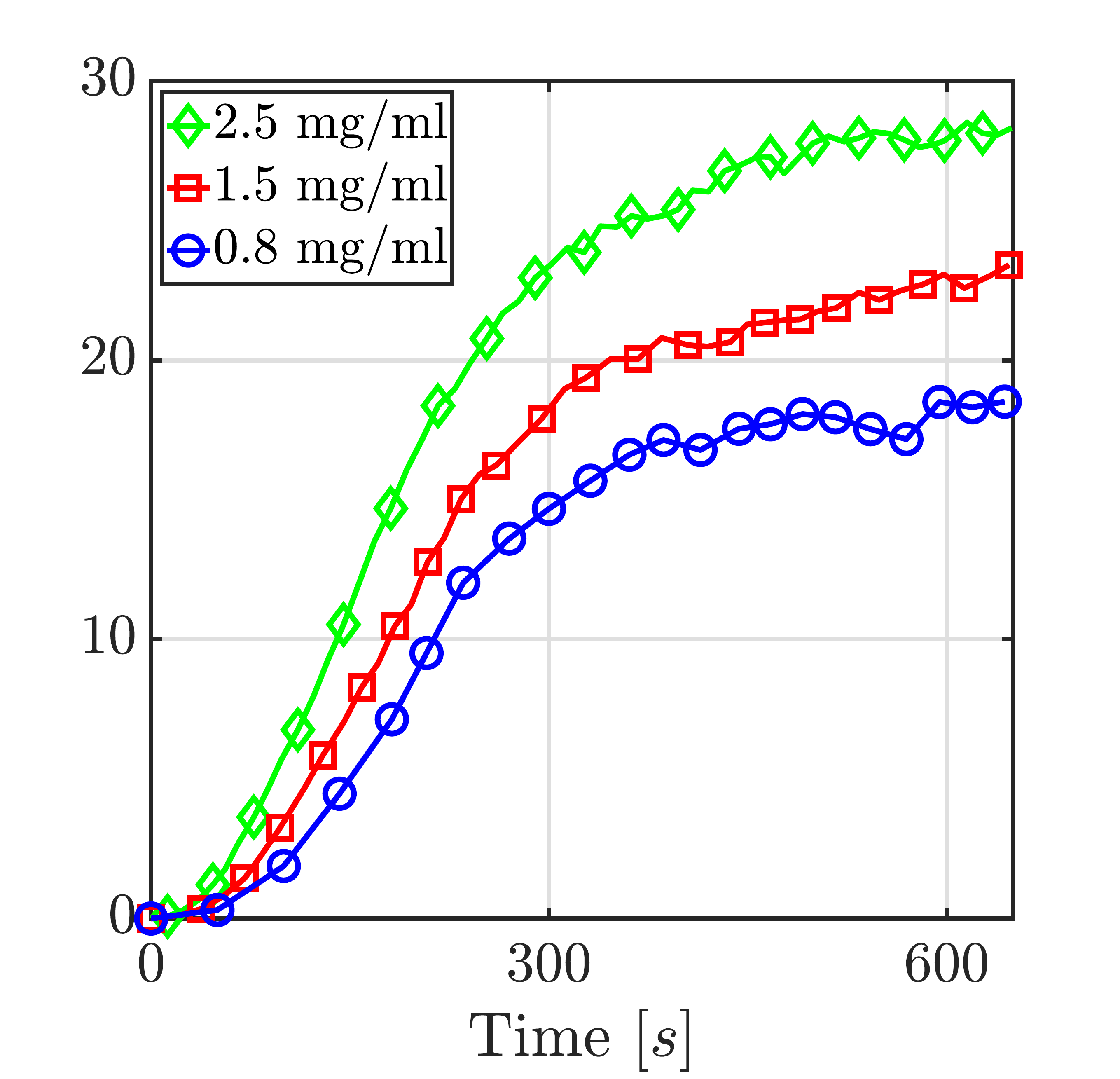}
\end{minipage}
\hfill
\begin{minipage}[b]{0.32\linewidth}
	\centering
	\includegraphics[width=1.0\linewidth]{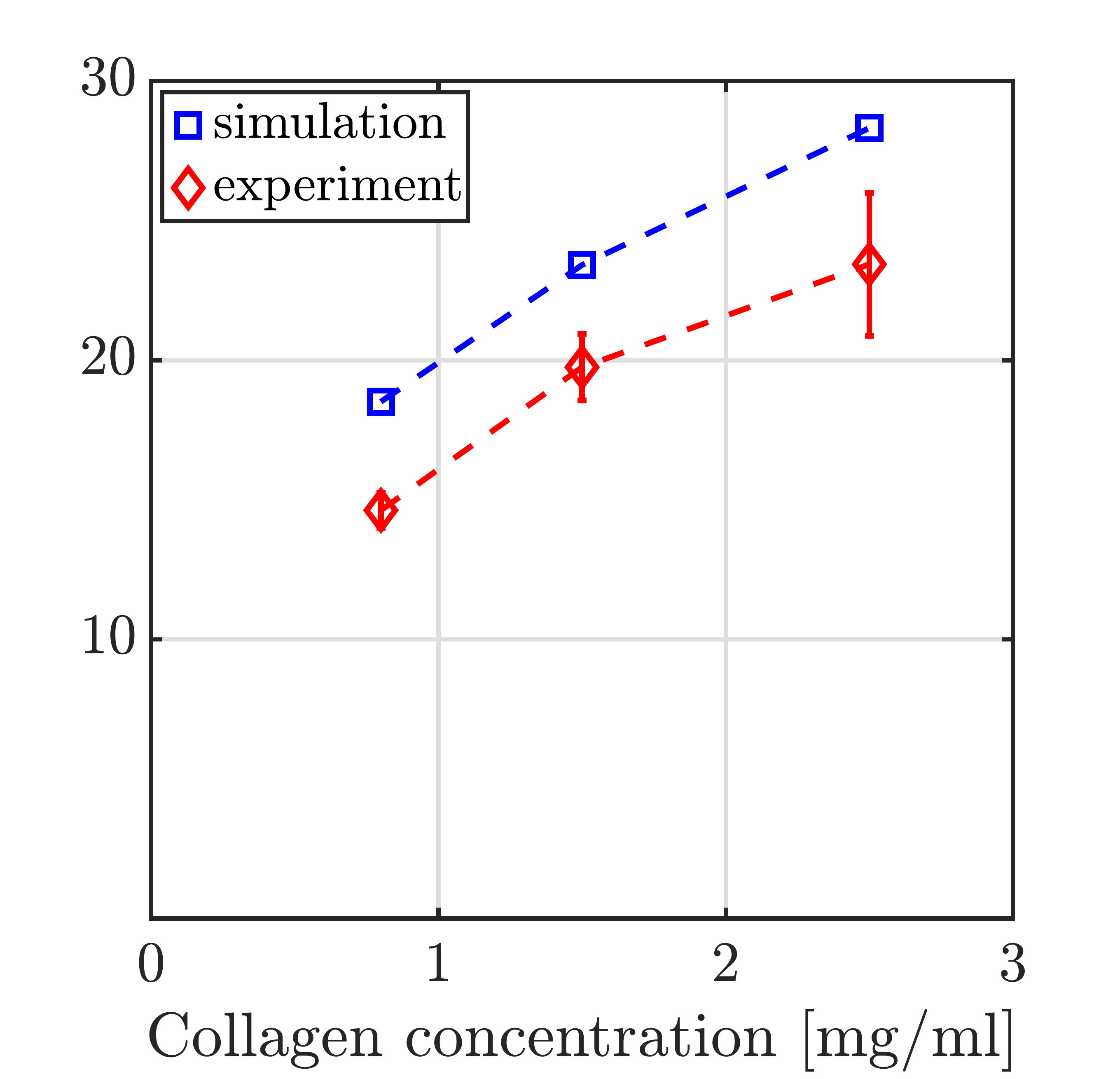}
\end{minipage}
\hfill
\begin{minipage}[b]{0.0\linewidth}
\end{minipage}
\hfill
\caption{Mechanical homeostasis for a cell concentration of $0.5 \cdot 10^6\ \text{cells/ml}$ and different collagen concentrations in (A) experiments \citep{Eichinger2020} and (B) our simulations. (C) In both cases the relation between homeostatic first Piola-Kirchoff stress (last data points were taken respectively) and collagen concentration is approximately linear.}
\label{fig:collagen_concentration}
\end{figure*}

It is well-known that interactions between cells and their environment crucially depend on the stiffness of the environment. This holds in particular for the proliferation, survival, migration, and differentiation of cells \citep{Wang2012a,Nguyen2018,Balcioglu2020}. A simple way of testing the impact of stiffness on cellular behavior in tissue culture studies is to change the collagen concentration of the tested gels \citep{Hall2016,Joshi2018,Alcaraz2011,Miroshnikova2011}. As shown in Fig. \ref{fig:collagen_concentration} A, tissue culture studies with a cell density of $0.5 \cdot 10^6\ \text{cells/ml}$ revealed that the cell-mediated first Piola-Kirchhoff stress increases in collagen gels with the collagen concentration \citep{Eichinger2020,Delvoye1991}. This behavior is both qualitatively and quantitatively reproduced well by our simulations as shown in Fig. \ref{fig:collagen_concentration} B. Interestingly, both experiments and simulations exhibit a nearly linear relation (with a slope of $\sim 9/2$) between collagen concentration and the homeostatic stress (Fig. \ref{fig:collagen_concentration} C). Moreover, the slope of the increase of stress up to the homeostatic stress was largely independent of the collagen concentration compared to the cell density in both the experiments and our simulations. We know from our simulations that an increased fiber density in case of higher collagen concentrations in combination with a constant distance between integrin binding spots on fibers of $50nm$ \citep{Lopez-Garcia2010} leads to more cell-matrix links per cell over time (data not shown) even when only the mechanisms presented in Section \ref{subsec:cell_ecm} are considered. If one assumes that cells stress fiber after fiber up to a certain level, this process takes longer if more fibers are present and can explain the observed nearly linear relationship between homeostatic stress and collagen concentration as well as the similar initial slope for all three collagen concentrations.

\textbf{\subsection{Residual matrix tension}}

\begin{figure*}[htbp] 
\begin{minipage}[c]{0.0\linewidth}
\end{minipage}
\hfill
\\
\begin{minipage}[c]{0.0\linewidth}
\end{minipage}
\hfill
\begin{minipage}[c]{0.60\linewidth}
A
\end{minipage}
\hfill
\begin{minipage}[c]{0.38\linewidth}
B
\end{minipage}
\hfill
\begin{minipage}[c]{0.0\linewidth}
\end{minipage}
\hfill
\\
\begin{minipage}[c]{0.0\linewidth}
\end{minipage}
\hfill
\begin{minipage}[c]{0.6\linewidth}
	\centering
	\includegraphics[width=1.0\linewidth]{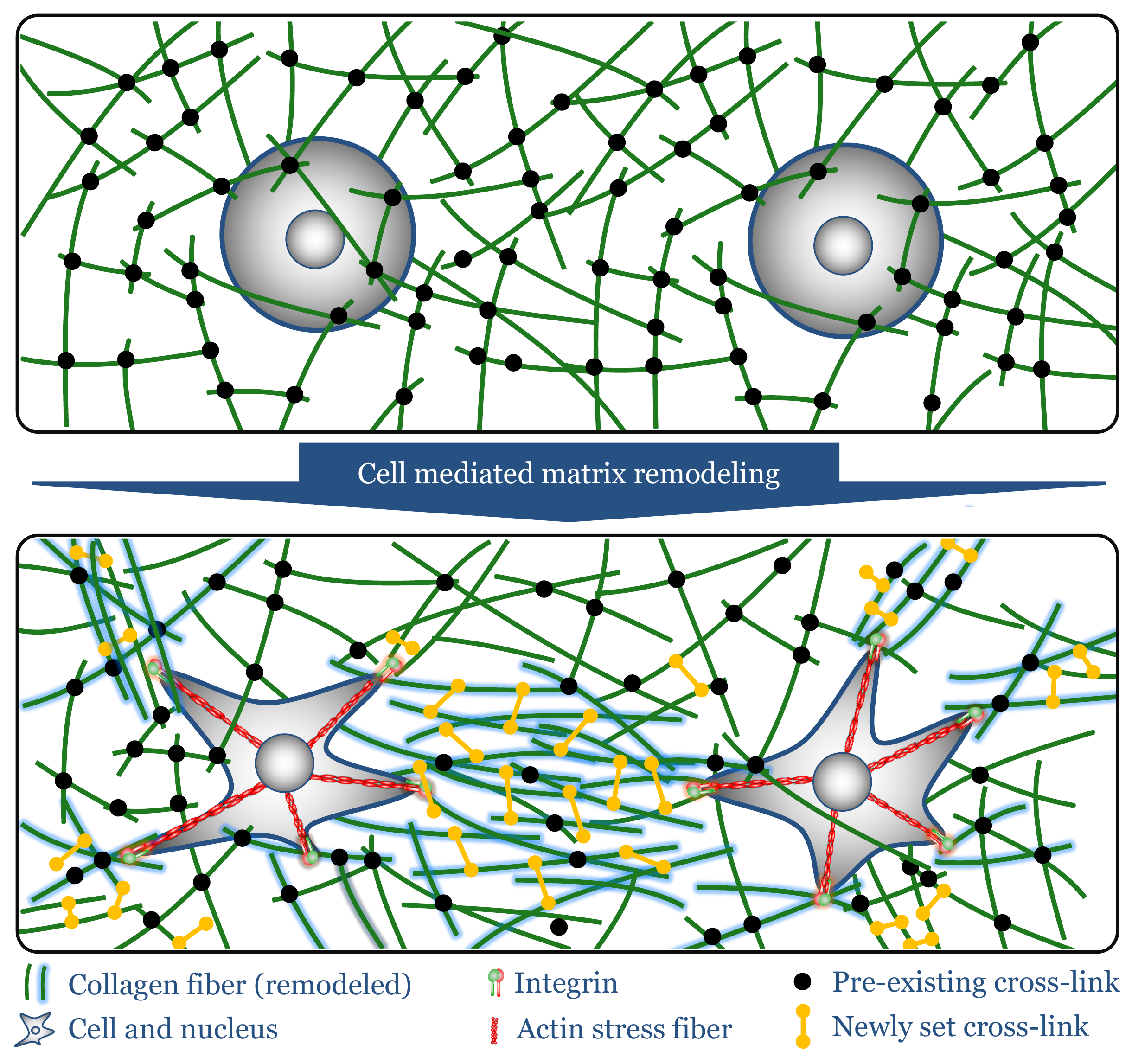}
\end{minipage}
\hfill
\begin{minipage}[c]{0.38\linewidth}
	\centering
	\includegraphics[width=1.0\linewidth]{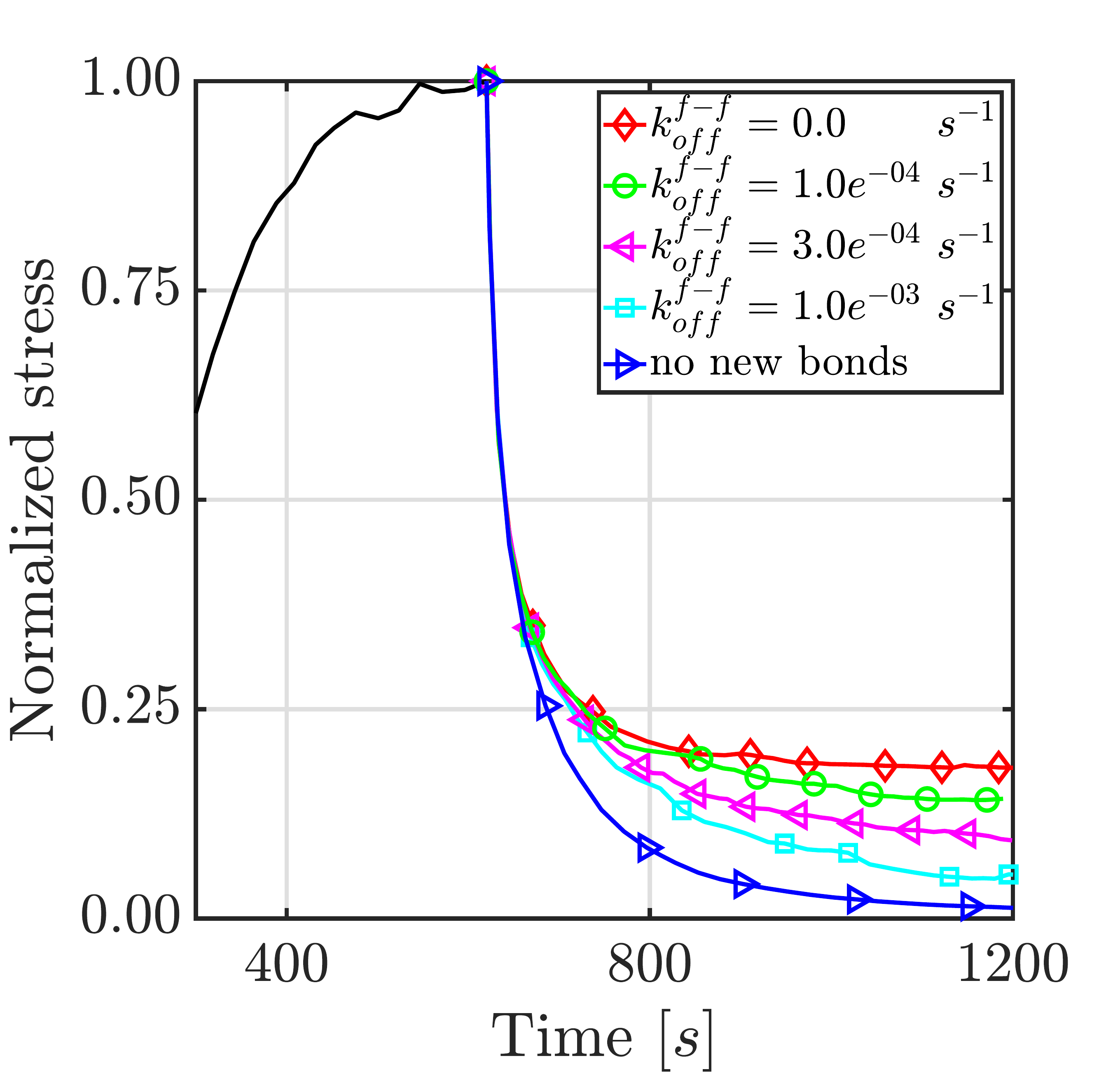}
\end{minipage}
\hfill
\begin{minipage}[c]{0.0\linewidth}
\end{minipage}
\hfill
\caption{(A) Cells actively remodel their surroundings, reorganizing the network and establishing new cross-links between its fibers. This way, cell-mediated tension can be entrenched in the network. (B) When removing active cellular forces suddenly, the matrix tension quickly drops. However, if cells have entrenched their reorganization of the network structure by permanent (covalent) cross-links (with $k^{f-f}_{off}=0.0$), a residual tension persists in the network. By setting transient cross-links with a sufficiently low off-rate, the cells can ensure an RMT at least over the periods considered.}
\label{fig:fiber_to_fiber_crosslinking}       
\end{figure*}

Mechanical homeostasis in soft tissues is closely lin\-ked to growth (changes in mass) and re\-mo\-deling (chan\-ges in microstructure) \citep{Cyron2017a}. In particular, a reorganization of the microstructure of tissues includes a change in the mechanical links between tissue fibers and of the constituent-specific natural (stress-free) configurations. Experimental studies have revealed that remodeling of collagen gels induced by cellular forces is time-dependent and inelastic \citep{Kim2017,Ban2018a}. Recent computational work suggested that the inelastic nature of cell-mediated remodeling is induced by force-dependent breaking of weak inter-fiber connections followed by the formation of bonds in new configurations leading to altered connections between tissue fibers \citep{Kim2017,Ban2019,Nam2016b,Cao2017} (Fig. \ref{fig:fiber_to_fiber_crosslinking} A). This implies that after cell-mediated remodeling, a part of the matrix tension remains in the tissue even after the elimination of all active cellular forces (e.g. by disrupting the actomyosin apparatus via addition of cytochalasin D or by cell lysis). This part is often referred to as residual matrix tension (RMT) \citep{Simon2014,Marenzana2006}. 

To date, our quantitative understanding of how an altered state of the matrix is entrenched during remodeling and how RMT develops is limited. Even the exact kind of cross-linking which occurs when matrix tension is entrenched is unknown. An inelastic change of the stress-free configuration of the tissue could emerge from newly formed, transient bonds between collagen fibers (such as hydrogen bonds or van der Waals forces) as a result of fiber accumulation in the surroundings of contractile cells \citep{Kim2017,Ban2018a}. However, RMT could also be entrenched by cells producing covalent cross-links via the actions of tissue transglutaminase or lysyl oxidase, which can also form new bonds between deformed matrix fibers. The impact of these enzymes on matrix remodeling has been shown experimentally in free-floating collagen gels \citep{Simon2014}. To study RMT, we simulated the experimental protocol presented in \cite{Marenzana2006} and eliminated active cellular forces from the simulated system in the homeostatic state by dissolving all existing cell-ECM bonds at a certain time (by setting $k^{c-f}_{on} =0$, which led to a rapid dissolution of the remaining bonds). We then tracked tension over time in the RVE. 

We first studied RMT in a purely covalently cross-linked network, implying that all existing bonds between fibers remained stable and no new bonds were formed during the simulation. After deactivating  active cellular forces, we observed a (viscoelastic) decline of tension to zero in the RVE (Fig. \ref{fig:fiber_to_fiber_crosslinking} B bottom curve). This finding suggested that networks that lack the ability to form new, at least temporary stable, bonds cannot entrench a residual tension in the matrix, which was however shown in the aforementioned experimental studies \citep{Simon2014,Marenzana2006}.

In a second step, transient linkers (which could, for example, be interpreted as un-bonded, freely floating collagen molecules or hydrogen bonds) were allowed to form between fiber-to-fiber binding spots with a certain on-rate $k^{f-f}_{on}$; they were able to be dissolved with a certain off-rate $k^{f-f}_{off}$. If two binding spots resided at some point in close proximity to two nearby fibers, a new, initially tension-free bond was formed according to Eq. \eqref{eqn:pon}. We found that introduction of newly formed, transient bonds enables the entrenchment of matrix remodeling and thus some RMT (Fig. \ref{fig:fiber_to_fiber_crosslinking} B, $k^{f-f}_{off}= 1.0e^{-04}\ s^{-1}$, $k^{f-f}_{off}= 3.0e^{-04}\ s^{-1}$, $k^{f-f}_{off}= 1.0e^{-03}\ s^{-1}$) at least for a prolonged period. The transient nature of the cross-links between the fibers resulted, however, in a slow decrease of RMT over time. This decrease happened faster, the higher the off-rate $k^{f-f}_{off}$ (Fig. \ref{fig:fiber_to_fiber_crosslinking} B). If $k^{f-f}_{off}$ was chosen above a certain threshold, we did not observe any RMT.

In a third study, we allowed covalent cross-linker molecules to form between two nearby collagen fibers when they were within a certain distance to each other and Eq. \eqref{eqn:pon} was fulfilled. By setting $k^{f-f}_{off} = 0$, a newly set bond could not be dissolved and was therefore covalent (permanent). In this case, we observed a substantial RMT that apparently did not decrease over time (Fig. \ref{fig:fiber_to_fiber_crosslinking} B, $k^{f-f}_{off}=0.0\ s^{-1}$). 

It thus appears that both transient and covalent cross-links play roles in inelastic matrix remodeling. Our study suggests that RMT crucially depends on the ability of cells to entrench the deformation they impose on their neighborhood by covalent, permanent cross-links. Such a permanent entrenchment appears energetically favorable because it releases cells from the necessity of maintaining matrix tension over prolonged periods by active contractile forces, which consume considerable energy.\\

\section{Conclusion}
\label{sec:conclusion}

To date, our understanding of the governing principles of mechanical homeostasis in soft tissues on short time spans especially on the scale of individual cells remains limited \citep{Eichinger2020b}. To address some of the many open questions in this area, we developed a novel computational framework for modeling cell-ECM interactions in three-dimensional RVEs of soft tissues. Our computational framework generates random fiber networks whose geometric characteristics resemble those of actual collagen type I gels, that is, they exhibit a similar distribution of valency, free-fiber length, and orientation correlation (direction cosine) between adjacent fibers. These microstructural characteristics have been shown to be the primary determinants of the mechanical properties of fiber networks \citep{Davoodi-Kermani2021}. To model the mechanics of the collagen fibers in the network, our framework discretizes these fibers with geometrically exact nonlinear beam finite elements, which were shown in Section \ref{subsec:uniaxial_test} to reproduce the elastic properties of collagen fiber networks. Our framework enables efficient parallel computing and can thus be used to simulate RVEs of tissues with realistic collagen concentrations and cell densities. 

The physical interactions of cells with surrounding fibers through stress fibers in the cytoskeleton and transmembrane proteins (integrins) are modeled by contractile elastic springs whose binding and unbinding dynamics closely resemble the situation in focal adhesions. We used the non-trivial, experimentally determined relations of both cell density and collagen concentration to the homeostatic stress to show that the mechanisms accounted for in our computational framework are sufficient to capture theses relationships. We also demonstrated how our framework can help to (quantitatively) examine the micromechanical foundations of inelastic cell-mediated matrix remodeling and RMT, which persists in the tissue even after active cellular forces have been removed.

Despite its advantages and broad experimental foundation, the proposed computational framework has some limitations that remain to be addressed. First, our model does not yet capture mass turnover, that is, the deposition and degradation of fibers, which are assumed to be crucial for mechanical homeostasis on long time scales \citep{Cyron2017a,Humphrey2002,Goriely2010,Cyron2016,Braeu2017}. Moreover, it models integrins but not other proteins playing a key role in the interactions between cells and surrounding matrix such as talin and vinculin  \citep{Ziegler2008,Das2014,Yao2014,Yao2016,Austen2015,Truong2015,Davidson2015,Elosegui-Artola2016,Ringer2017,Grashoff2010,Carisey2013,Dumbauld2013}. Also the model of cellular contractility is simplistic and should be endowed with additional biological details \citep{Mogilner2003,Murtada2010,Murtada2012}.  Finally, we did not consider contact forces between fibers or between cells and fibers (assuming that cells and fibers mainly interact via integrins). While this reduces the computational cost substantially, a comprehensive incorporation of contact mechanics could also help to make our computational framework more realistic. 

An important field of application for our computational framework will be \textit{in silico} studies in which one can test step by step which additional features have to be incorporated in the framework to capture more and more phenomena observed \textit{in vitro} and \textit{in vivo}. Like this, it may contribute to uncover the micromechanical foundations of mechanical homeostasis on the level of individual cells and fibers and help to understand how these microscopic processes lead to what we call mechanical homeostasis on the macroscale.

\newpage

\clearpage 
\beginappendix
\section*{Appendix}
\label{sec:suplementary}

\section{Construction of random fiber networks by simulated annealing}
\label{sec:appendix_simanneal}

\begin{figure*}[htbp] 
\begin{minipage}[b]{0.0\linewidth}
\end{minipage}
\hfill
\begin{minipage}[b]{0.0\linewidth}
\end{minipage}
\hfill
\begin{minipage}[b]{0.32\linewidth}
A
\end{minipage}
\hfill
\begin{minipage}[b]{0.32\linewidth}
B
\end{minipage}
\hfill
\begin{minipage}[b]{0.32\linewidth}
C
\end{minipage}
\hfill
\begin{minipage}[b]{0.0\linewidth}
\end{minipage}
\hfill
\\
\begin{minipage}[b]{0.0\linewidth}
\end{minipage}
\hfill
\begin{minipage}[b]{0.32\linewidth}
	\centering
	\includegraphics[width=1.0\linewidth]{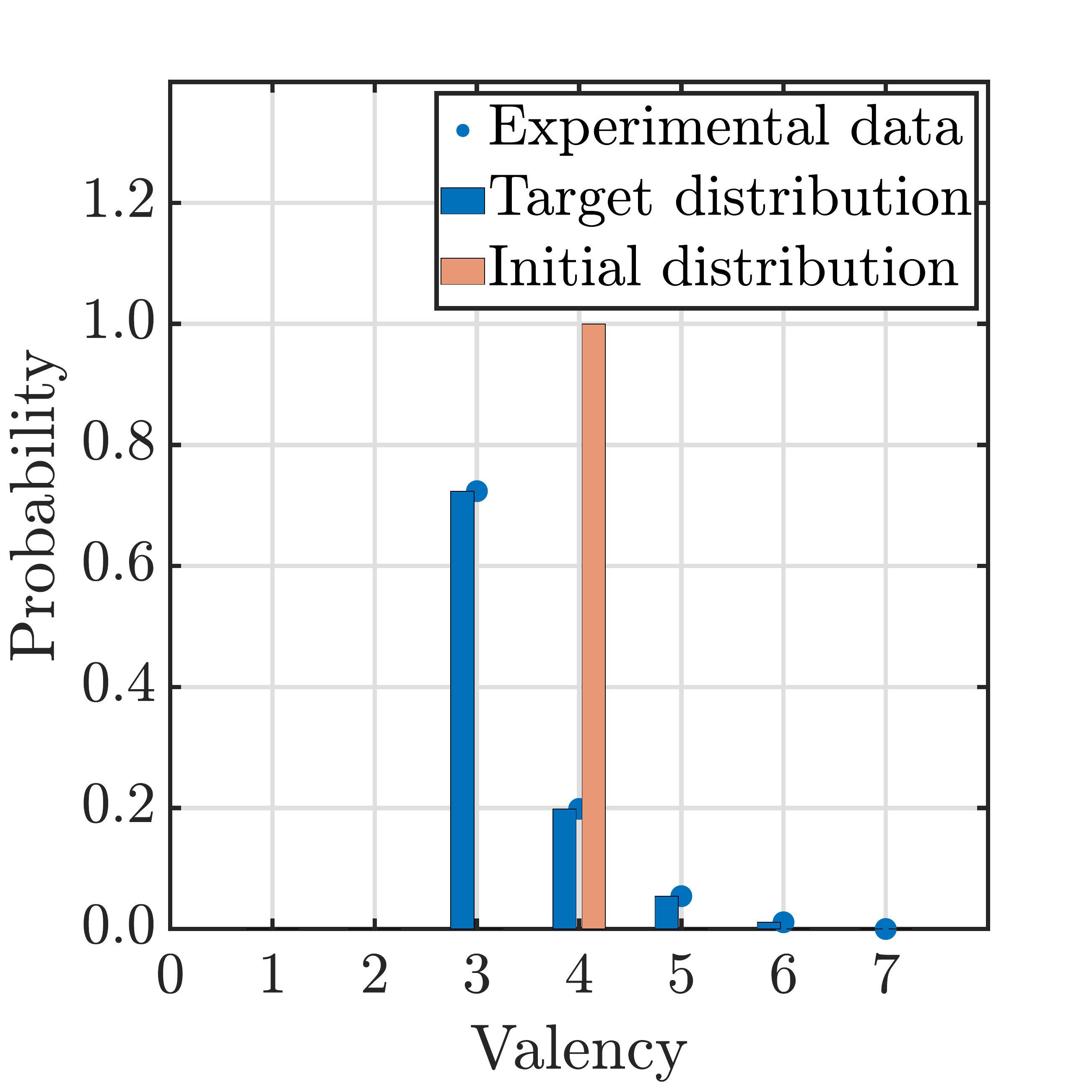}
\end{minipage}
\hfill
\begin{minipage}[b]{0.32\linewidth}
	\centering
	\includegraphics[width=1.0\linewidth]{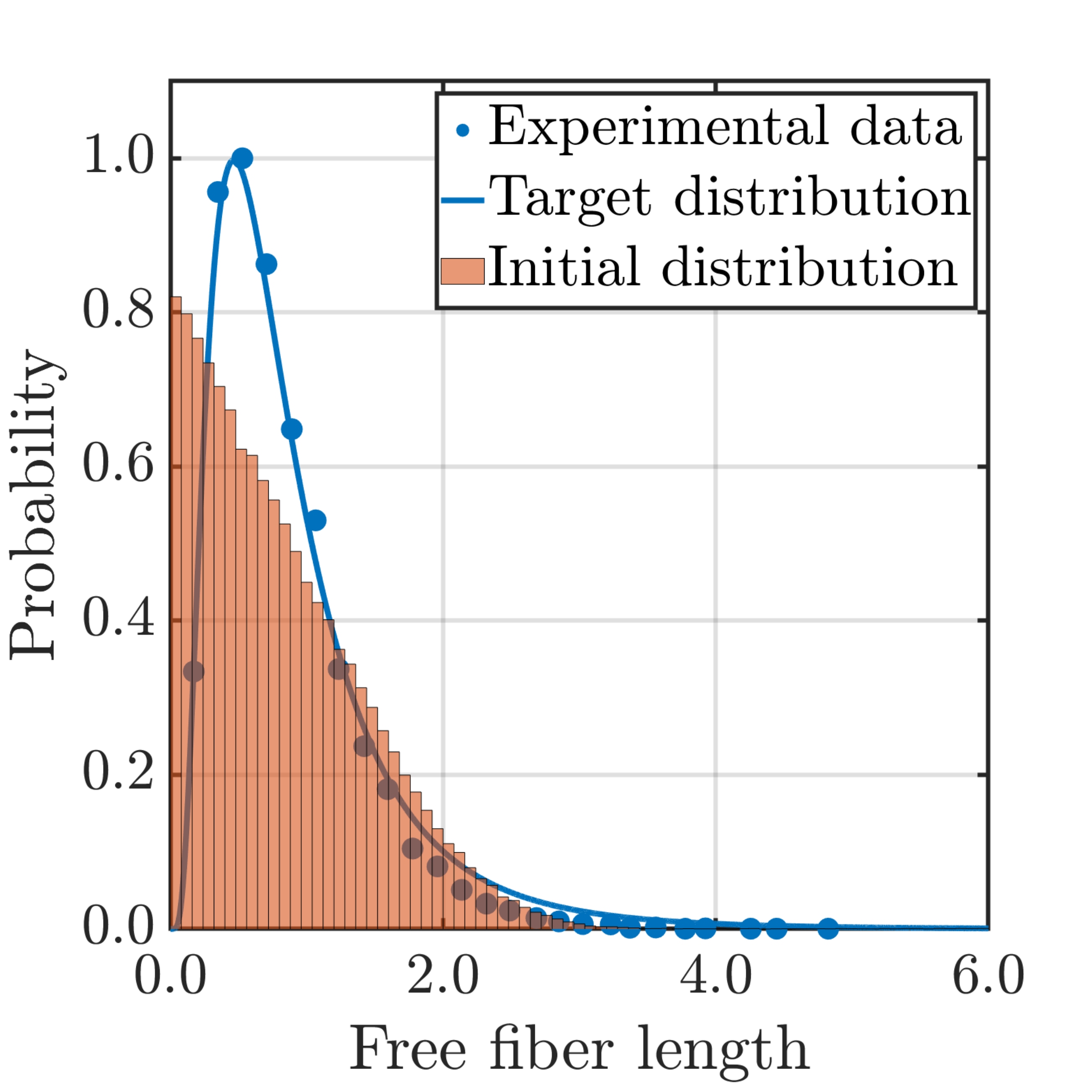}
\end{minipage}
\hfill
\begin{minipage}[b]{0.32\linewidth}
	\centering
	\includegraphics[width=1.0\linewidth]{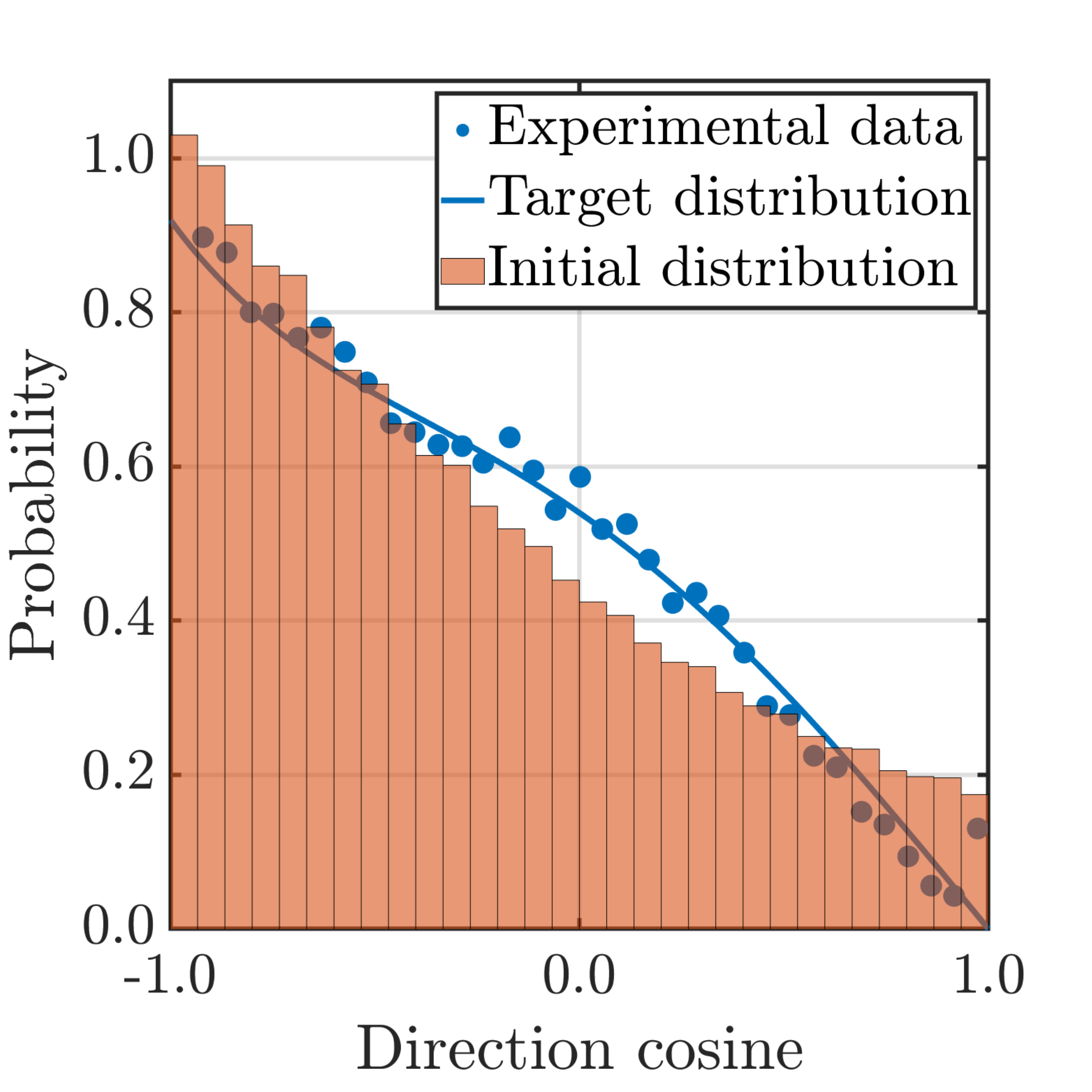}
\end{minipage}
\hfill
\caption{Random initial descriptor distributions in a network generated by Voronoi tesselation vs. target distributions fitted by \cite{Lindstrom2010} and \cite{Nan2018} to experimental data (collagen concentration $2.5 mg/ml$). Simulated annealing alters the initial network until its descriptor distributions match the required target distributions.}
\label{fig:sim_annealing_results_initial}
\end{figure*}

In this appendix, we present the computational details of the algorithm we used for constructing network RVEs as an input for our simulations. Our algorithm closely follows the approach of \cite{Lindstrom2010}, using the stochastic optimization method of simulated annealing for constructing random heterogeneous media introduced by \cite{Yeong1998}. Thereby, one assumes that the geometry of a fiber network can be characterized by some descriptors $\bm{x}_i$, with $i \in \{l, c\}$, in our case representing the fiber length and the direction cosine, respectively. These descriptors can be understood as random variables taking on specific values at certain nodes or fibers and characterize the network microstructure. The descriptors are assumed to follow some statistical distribution $P^{i}(x_i)$ across the different fibers and nodes. These distributions can be determined, for example, from confocal microscopy images of real networks, see also Fig. \ref{fig:sim_annealing_results_initial}. According to \cite{Lindstrom2010}, this yields for collagen type-I networks 
\begin{align}
P^l(l) &= \frac{1}{l\sigma \sqrt{2\pi}} \exp\left( -\frac{[\mu - \ln(l)]^2}{2\sigma^2} \right),
\label{eqn:length_pdf}
\end{align}
where $l$ denotes the fiber length normalized by $(N/V_{RVE})^{\frac{1}{3}}$, with $V_{RVE}$ being the volume of the RVE and $N$ representing the total number of network nodes in it. The parameters $\sigma$ and $\mu$ denote a standard deviation and mean value that may vary from network to network. Typical parameters are given in Table \ref{table:reconstruction_mat_params}. The cumulative probability distribution associated with $P^l(l)$ is given by
\begin{align}
C^l(l) &= \frac{1}{2} + \frac{1}{2} \erf\left(\frac{\ln(x) - \mu}{\sqrt{2}\sigma} \right)
\label{eqn:length_cdf}
\end{align}
and will be used below in Eq. \eqref{eqn:sim_ann_binning}. 

The distribution of the direction cosine $\beta$ of fibers adjacent to the same node has been described by \cite{Lindstrom2010} by a truncated power series
\begin{align}
P^c(\beta) &= \sum_{k=1}^{3} b_k\left( 1-\beta\right) ^{2k-1},
\label{eqn:cosine_pdf}
\end{align}
with the associated cumulative distribution function
\begin{align}
C^c(\beta) &= 1 + \sum_{k=1}^{3} -\frac{b_k}{2k} \left( 1-\beta\right) ^{2k-1}.
\label{eqn:cosine_cdf}
\end{align}
Again, typical values for the parameters $b_k$ are given in Table \ref{table:reconstruction_mat_params}. To describe the valency distribution of the networks, we relied on the data reported in \cite{Nan2018}.

\begin{table*}
\centering
\caption{Parameters for length, valency and cosine distribution functions according to \cite{Lindstrom2010} and parameters used for simulated annealing process}
\label{table:reconstruction_mat_params}
\begin{tabular}{clr}
Parameter & Description & Value [-] \\
\midrule
$w_{l}$ & weight for free-fiber length distribution in Eq. \eqref{eqn:energy} & $1.0 $ \\
$w_{c}$ & weight for direction cosine distribution in Eq. \eqref{eqn:energy} & $1.0 $ \\
$\mu$  & mean in Eq. \eqref{eqn:length_pdf} and \eqref{eqn:length_cdf} & $-0.3000$  \\
$\sigma$ & standard deviation in Eq. \eqref{eqn:length_pdf} and \eqref{eqn:length_cdf} & $0.6008$  \\
$b_1$ & parameter for truncated power series in Eq. \eqref{eqn:cosine_pdf} and \eqref{eqn:cosine_cdf} & $0.6467$ \\
$b_2$ & parameter for truncated power series in Eq. \eqref{eqn:cosine_pdf} and \eqref{eqn:cosine_cdf} & $-0.1267$ \\
$b_3$ & parameter for truncated power series in Eq. \eqref{eqn:cosine_pdf} and \eqref{eqn:cosine_cdf} & $0.0200 $ \\
$b_{l}$ & number of bins for free-fiber length distribution in Eq. \eqref{eqn:sim_ann_binning} & $1000 $ \\
$b_{c}$ & number of bins for direction cosine distribution in Eq. \eqref{eqn:sim_ann_binning} & $1000 $ \\
$T_{0}$ & initial temperature & $0.05 $ \\
- & resulting average node valency of constructed networks  & $3.3 $ \\
\bottomrule
\end{tabular}
\end{table*}

Our target was to construct artificial random fiber networks as an input for our simulations whose descriptor distributions matched the ones defined above. To this end, we started from some random initial network. This network was then evolved in a number of discrete random steps according to the concept of simulated annealing \citep{Kirkpatrick1983}, until the descriptor distributions matched the desired target distributions.

To define the random initial configuration, we started by generating networks based on three-dimensional Vo\-ro\-noi tesselations \citep{Rycroft2009} with periodic boundary conditions applied in all directions. Subsequently, we randomly removed and added fibers until the valency distribution matched its target distribution. Only then we started the actual simulated annealing, where only fiber length and direction cosine distributions still had to be matched to their target distributions. The simulated annealing was performed following the concept introduced by \cite{Kirkpatrick1983}. The idea is to iteratively select random nodes in the network and apply random displacements to them (Fig. \ref{fig:network_reconstruction} B). Like this, the length of all fibers attached to the respective node and the angles between these fibers change. Importantly, only movements of nodes are accepted which do not lead to fiber lengths larger than one third of the smallest edge length of the RVE to ensure that it stays \textit{representative}. Note, that a movement of a node does not affect its connectivity, which ensures that the initially created valency distribution remains unaffected during the whole simulated annealing.

For stochastic optimization according to the simulated annealing concept, it is helpful to define an objective (energy-type) function $E$ 
\begin{align}
E &= w_l \cdot E^l + w_c \cdot E^c .
\label{eqn:energy}
\end{align}
where the $E^l$ and $E^c$ become minimal if the length and direction cosine distribution exactly match their target distributions and where the $w_{i} > 0$ are weights that can be adapted to tune the importance of a specific distribution function. Having defined the objective function $E$, simulated annealing can be understood as a stochastic minimization of $E$. Once the minimum is found, $E^l$ and $E^c$ must be minimal and thus the length and direction cosine distributions match their target distributions. To perform a stochastic minimization of $E$, a Metropolis algorithm is applied during the simulated annealing. It consists of a sequence of random steps. For each of these steps the associated change of $E$ is computed, that is, $\Delta E$. Then, the step is actually performed only with a likelihood

\begin{align}
p_{accept}(\Delta E)&= 
\begin{cases}
1, & \Delta E \leq 0 \\
\exp(-\frac{\Delta E}{T}), & \Delta E > 0,
\end{cases}
\label{eqn:delta_E}
\end{align}
where $T$ denotes a temperature-like parameter. In our simulated annealing, we slowly decreased $T$ as the number random steps increased, using at the annealing step $k$ the value $T = 0.95^k \cdot T_0$ (according to \cite{Nan2018}). We chose $T_0$ such that the probability for accepting a random step with $\Delta E > 0$ was approximately $0.5$ in the beginning. In practice, the simulated annealing was stopped if either the total energy of the system was below a predefined threshold or a maximal number of iterations was reached. 

\paragraph{\textbf{Remark A1:}} It is worth noting, that for constructing RVEs with different collagen concentrations, we assumed the same target distributions for the valency, direction cosine and normalized fiber length. Only the normalization factor of the fiber length was changed. Moreover, an increased collagen concentration automatically also implies a higher number $N$ of network nodes in the RVE. 

\paragraph{\textbf{Remark A2:}} While there exists a variety of simple and obvious choices for the $E^i$ in \eqref{eqn:energy}, these mostly suffer from a computational cost on the order of $\mathcal{O}(n_i)$ with $n_i$ being the number of instances of a descriptor. This makes the generation of large random networks practically infeasible. To overcome this problem, we adopted the idea of \cite{Lindstrom2010} to use a binning algorithm and define the $E^i$ as Cramer-von Mises test statistics, which reduces the computational cost to the order of $\mathcal{O}(b)$ with $b$ being the number of bins. To this end, we divided the range of $\bm{x_i}$ in $b_i$ disjoint intervals (bins) and assigned each instance of a descriptor at a fiber or node of a random network to its associated bin. The resulting histogram is a discrete approximation of $P^i(x_i)$. The center of the $j$-th bin is denoted by $x_{ij}$. The number of instances of descriptor $x_i$ assigned to the $j$-th bin is $m_{ij}$. The Cramer-von Mises test statistics can then be computed as
\begin{equation}
\begin{split}
E^i = \frac{1}{n_i^2}\sum_{j=1}^{b_i} m_{ij} \left[\frac{1}{6}(m_{ij}+1)(6S_{ij}+2m_{ij}+1)+S_{ij}^2 \right] 
\end{split}
\label{eqn:sim_ann_binning}
\end{equation}
with $S_{ij} = M_{i(j-1)} - nC^i(x_{ij}) - \frac{1}{2}$, $M_{ij} = \sum_{k=1}^{j}m_{ik}$, and $C^i$ the cumulative distribution of $x_i$.

\section{Boundary conditions}
\label{sec:boundary_conditions}

\begin{figure*}[]
\begin{center}
 \includegraphics[width=1.0\textwidth]{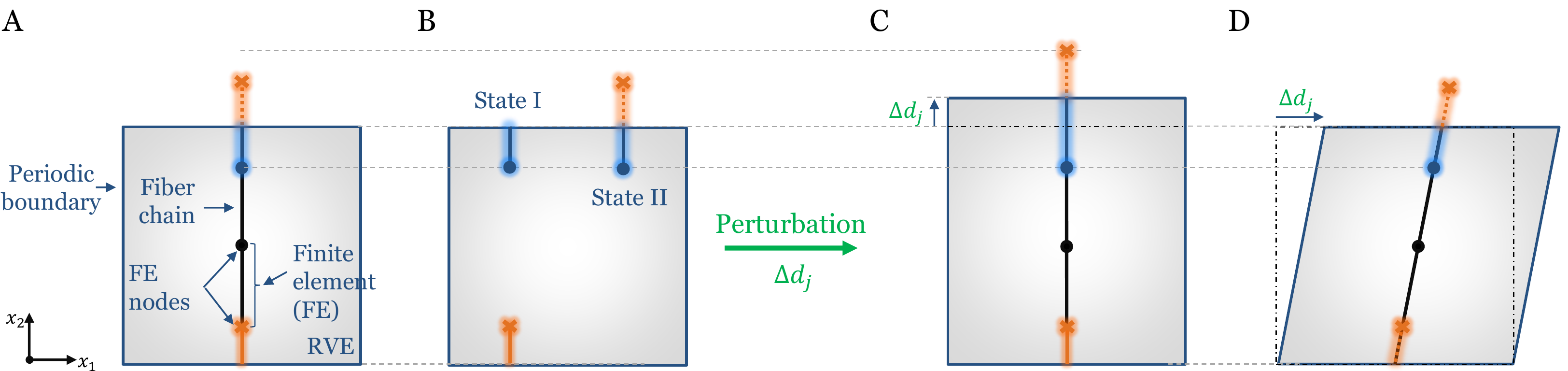}
\end{center}  
\caption{Illustration of periodic boundary conditions using the example of a single fiber in the network: (A)  any fraction of an element sticking out of a periodic boundary must have a counterpart entering at the opposite side; (B) both element fractions together define what is physically present within the RVE (state I). To compute strains and stresses in both element fractions, it is convenient to use a fictitious state II (shifted rightwards in the figure for illustration purposes only), which represents the part of the cut element within the simulated RVE and the part located in an adjacent domain periodically continuing the RVE; (C) application of fully periodic normal strain boundary condition in vertical direction; (D) application of fully periodic shear strain boundary condition in the drawing plane.}
\label{fig:pbc}       
\end{figure*}

In this appendix, we briefly summarize how we applied fully periodic boundary conditions to our simulation domains. Let these domains be cuboids with edge length $L_i$ in the $i$-th coordinate direction. In a fully periodic network, the part of a fiber sticking out across one periodic boundary must have a counterpart entering the RVE at the opposite side (Fig. \ref{fig:pbc} A). One can interpret the element part sticking out of the domain at one boundary and the element part entering the RVE at the opposite boundary also as a fictitious single element (Fig. \ref{fig:pbc} B, state II) cut into two parts (Fig. \ref{fig:pbc} B, state I). Thereby, state I as delineated in Fig. \ref{fig:pbc} B can be used to evaluate interactions with other fibers or cells, and state II for evaluating strains and stresses on element level. If the element is cutting through a boundary in the i-th coordinate direction, the i-th coordinate of the nodal positions in state I and state II are shifted by $L_i$ relative to each other. Importantly, only the translational degrees of freedom of the beam finite element nodes are affected by the periodic boundaries, rotational degrees of freedom remain unaffected. 

It is a major challenge to impose periodic Dirichlet boundary conditions on fiber networks in a manner that is fully periodic. Note that most of the literature \citep{Stein2011,Lee2014,Humphries2018,Abhilash2014,Burkel2018,Ban2018a,Liang2016a,Ban2019} bypasses this difficulty by fixing nodes on or close to the periodic boundary in a manner that actually unfortunately cannot ensure periodicity in a rigorous manner. To overcome this deficiency, we used the following approach. Dirichlet boundary conditions of RVE can be represented by normal or shear strains. These strains can be converted into a relative displacement of opposite periodic boundaries by components $\Delta d_j$ in the $j$-th coordinate direction. We accounted for this displacement by stretching (Fig. \ref{fig:pbc} C) or shearing (Fig. \ref{fig:pbc} D) the RVE as a whole. The nodal positions in state I and II were then no longer converted into each other in the above described simple manner, that is, by a relative shift by $L_i$ in the $i$-the direction. Rather all coordinates of the nodal positions were additionally shifted relative to each other by the components $\Delta d_j$. Note that this approach can account also for complex multi-axial loading by applying the described procedure at all periodic boundaries. Moreover, this approach can account also for large strains.

\section{Search algorithm and parallel computing}
\label{sec:par_search}

\begin{figure*}[htb!] 
\begin{center}
 \includegraphics[width=0.7\textwidth]{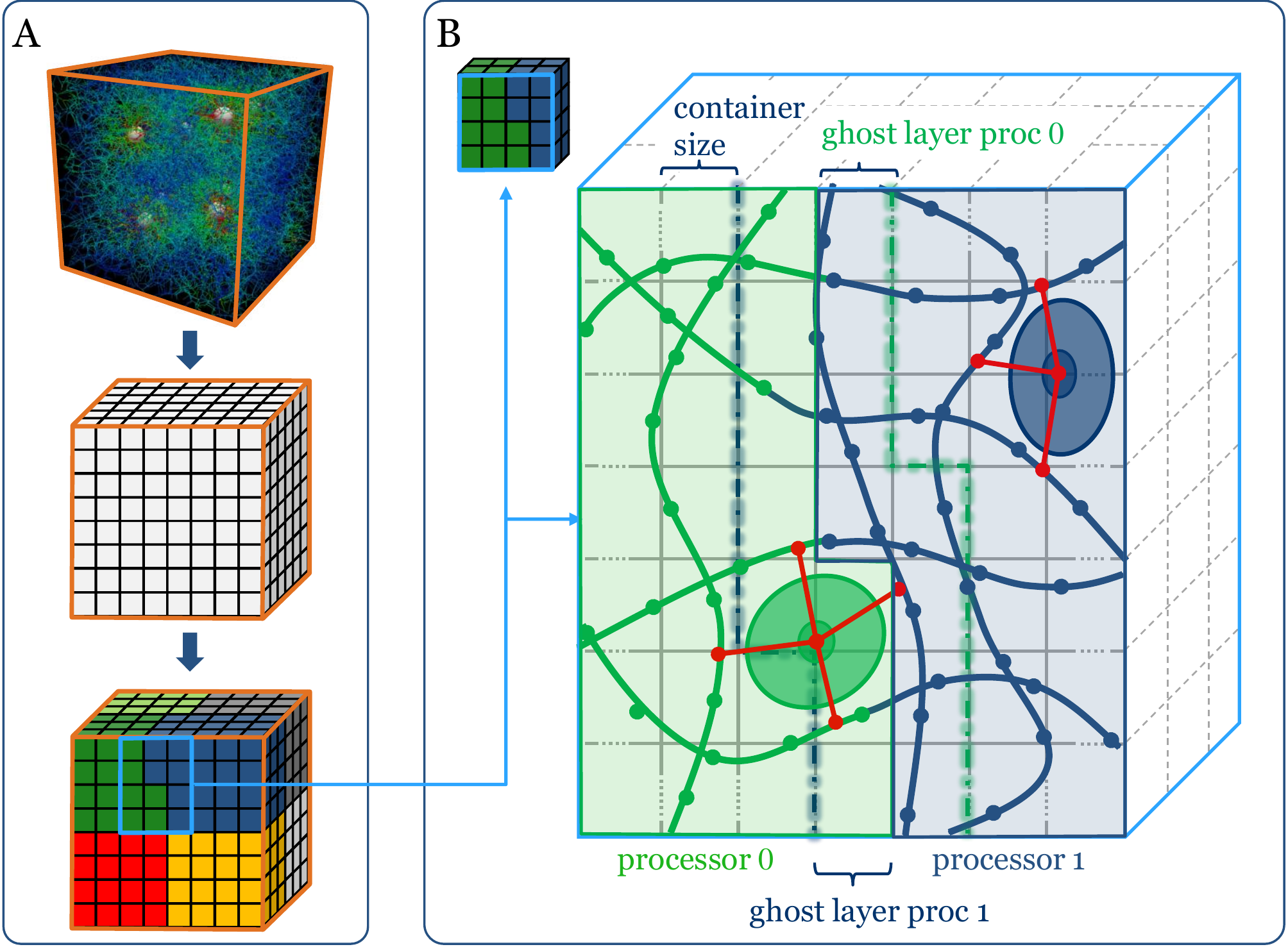} \\
\end{center}  
\caption{(A) Our computational domain (top) was divided into a large number of cubic containers (middle). Sets of numerous such containers (highlighted by different colors, bottom) were distributed to different processors. (B) All fibers discretized by beam finite elements as well as all cells were assigned to all containers with which they overlapped. Each processor was provided not only information about its own containers but also about a layer of ghost containers with whose elements the elements in its own container may interact.}
\label{fig:binning}
\end{figure*}

In this appendix we describe how we ensured efficient parallel computing for the presented modeling framework in our in-house finite element solver BACI \citep{Baci2020}. Parallelization of the finite element discretization of the fibers can be handled with standard libraries such as the Trilinos libraries that formed the basis of our in-house code. Therefore, we focus herein on the parallelization of cell-fiber interactions and chemical bonds between fibers. Both require search algorithms to identify cell-fiber or fiber-fiber pairs that may interact at a certain point in time. We implemented a search algorithm
based on a geometrical decomposition of the computational domain (RVE) in uniform cubic containers. For simplicity, these were aligned with the axes of our coordinate system (see Fig. \ref{fig:binning}). Cells and finite beam elements are assigned to all containers with which they overlap. We chose the minimal size of the containers such that all possible interaction partners were certainly located within one layer of neighboring containers. Hence, evaluating the possible interactions of a single cell or beam finite element simply required searching within one layer of containers around the containers to which the cell or element was assigned. 

The content of the containers had to be updated over time as cells and matrix fibers moved during our simulations. Depending on the time step size of our simulations, container size and effective physical interaction distance, it was feasible to update our containers only every $n$-th time step.

The potentially large domain considered in our simulations typically required a distribution of the above described containers on several processors. To this end, each processor was assigned a set of containers forming a connected sub-domain. In addition to these containers, each processor was also provided full information about one layer of so-called ghost containers surrounding its specific sub-domain (Fig. \ref{fig:binning} B). The computational cost of sharing the information about ghost containers was negligible compared to the the overall computational cost of our simulations. 

To always allow an effective search algorithm based on a rectangular Cartesian domain, we used
a coordinate transformation to the undeformed domain in case boundary conditions imposing a deformation of the computational domain.

It is worth mentioning that in our parallelization framework no data (except some uncritical parameters such as the current time step) need be stored fully redundantly on all processors, which would drastically limit the problem sizes.

\newpage
\section{Simulation parameters}
\label{subsec:simparams}

\begin{table*}[tbhp]
\centering
\caption{List of parameters and default values  of computational model}
\label{table:model_parameters}
\begin{tabular}{lllr}
Parameter & Description & Value & Reference \\
\midrule
$a_1$ & Integrin catch-slip bond parameter & $2.2$ & to fit data of \cite{Kong2009} \\
$b_1$ & Integrin catch-slip bond parameter & $29.9$ & to fit data of \cite{Kong2009} \\
$c_1$ & Integrin catch-slip bond parameter & $8.4$ & to fit data of \cite{Kong2009} \\
$a_1$ & Integrin catch-slip bond parameter & $1.2$ & to fit data of \cite{Kong2009} \\
$b_2$ & Integrin catch-slip bond parameter & $16.2$ & to fit data of \cite{Kong2009} \\
$c_3$ & Integrin catch-slip bond parameter & $37.8$ & to fit data of \cite{Kong2009} \\
$R$ & Cell radius & 12 $\mu m$ & typical value \\
$\Delta R$ & Linking range around cell & $\pm 3\mu m$ & - \\
$D_{f}$ & Diameter of collagen fibers & $180\ nm$ & \cite{VanDerRijt2006a} \\
$E_{f}$ & Young\textquotesingle s Modulus of collagen fibers & $1.1MPa$ & \cite{Jansen2018} \\
$\dot{c}$ & Contraction rate stress fiber & $0.1 \frac{\mu m}{s}$ & \makecell[r]{\cite{Choquet1997}\\\cite{Moore2010}} \\
$k_{B}T$ & Thermal energy & $4.28 \cdot 10^{-3}\ aJ$ & at 37\textcelsius \\
$L_i$ & RVE edge length in $i$-th coordinate direction & $245 \mu m$ & - \\
$k^{f-f}_{on}$ & Chemical association rate for fiber linker & $0.0001 s^{-1}$ & - \\
$k^{f-f}_{off}$ & Chemical dissociation rate for fiber linker& $0.0001 s^{-1}$ & - \\
$\Delta x$ & Bell parameter & $0.5 nm $ & - \\
$N_{FA,max}$ & Maximal number of focal adhesion per cell & $65 $ &\makecell[r]{ \cite{Horzum2014}\\ \cite{Kim2013}\\ \cite{Mason2019}} \\
$N_{i,FA,max}$ & Maximal number of integrins per focal adhesion & $1000$ & \makecell[r]{\cite{Wiseman2004}\\ \cite{Elosegui-Artola2014}} \\
$N_{i,ic,max}$ & Maximal number of integrins per cluster & $20$ &  \makecell[r]{\cite{Changede2015} \\ \cite{Cheng2020}} \\
$k^{c-f}_{on}$ & Chemical association rate for integrin & $0.1s^{-1}$ & slightly modified \cite{Elosegui-Artola2016} \\
$d^{i-f}$ & Distance between binding spots for integrin-fiber links &$ 50 nm$ & \cite{Lopez-Garcia2010} \\

\bottomrule
\end{tabular}
\end{table*}

\newpage

\end{document}